\documentclass[twocolumn]{aa}  
\usepackage{graphicx}
\usepackage{txfonts}
\usepackage{t1enc}
\usepackage{epsfig}
\usepackage{rotating}
\usepackage{verbatim}
\usepackage{arydshln}
\newcommand{\Tex}   {T_\mathrm{ex}}

\newcommand{\kms}   {km~s$^{-1}$}

\newcommand{\cmt}   {cm$^{-3}$}
\newcommand{\jpb}   {$\rm Jy~beam^{-1}$}    
\newcommand{\lo}    {$L_{\sun}$}
\newcommand{\mo}    {$M_{\sun}$}
\newcommand{\nh}    {NH$_3$}
\newcommand{\nth}   {N$_2$H$^+$}
\newcommand{\chtoh} {CH$_3$OH}
\newcommand{\water} {H$_2$O}

\newcommand{\et}    {et al.}
\newcommand{\eg}    {e.\,g.,}
\newcommand{\ie}    {i.\,e.,}
\newcommand{\hii}   {\ion{H}{ii}}
\newcommand{\uchii} {UC\ion{H}{ii}}

\newcommand{\phn}   {\phantom{0}}
\newcommand{\phnn}  {\phantom{0}\phantom{0}}

\newcommand{\phe}   {\phantom{$^\mathrm{c}$}}

\begin{document}
%

   \title{Three intermediate-mass YSOs with different properties
     emerging from the same natal cloud in IRAS\,00117+6412\thanks{The
       fits files for Figs.~\ref{fi_cont}, \ref{fn2hch}, and \ref{fi_12co_channel} are also
       available in electronic format at the CDS via anonymous ftp to
       cdsarc.u-strasbg.fr (130.79.128.5) or via
       http://cdsweb.u-strasbg.fr/cgi-bin/qcat?J/A+A/.}}

   \author{
         Aina Palau\inst{1}
          \and
         \'A. S\'anchez-Monge\inst{2}
   	  \and
   	 G. Busquet\inst{2}
	  \and
         R. Estalella\inst{2}
	  \and
         Q. Zhang\inst{3}
         \and
         P.~T.~P. Ho\inst{3,4}
          \and
          M.~T. Beltr\'an\inst{5}
	  \and
          H. Beuther\inst{6}
          }

   \institute{Centro de Astrobiolog\'{\i}a (INTA-CSIC),
              Laboratorio de Astrof\'{\i}sica Estelar y Exoplanetas, 
              LAEFF campus, P.O. Box 78, E-28691 Villanueva de la Ca\~nada, 
              Madrid, Spain\\
              \email{apalau@laeff.inta.es}
              \and
	      Departament d'Astronomia i Meteorologia (IEEC-UB),
	      Institut de Ci\`encies del Cosmos, Universitat de Barcelona,
	      Mart\'i i Franqu\`es, 1, E-08028 Barcelona, Spain
	      \and
	      Harvard-Smithsonian Center for Astrophysics, 60 Garden Street,
	      Cambridge, MA 02138, USA
	      \and
              Academia Sinica, Institute of Astronomy and Astrophysics, P.O. Box
              23-141, Taipei  106, Taiwan
	      \and
	      INAF, Osservatorio Astrofisico di Arcetri, Largo E. Fermi 5,
	      50125 Firenze, Italy
	      \and
	      Max-Planck-Institut for Astronomy, K$\rm \ddot{o}$nigstuhl 17, 69117
	      Heidelberg, Germany
             }

   \date{Received / Accepted}
   
   \authorrunning{Palau \et}
   \titlerunning{Three intermediate-mass YSOs in the making}

 
  \abstract
    {}
   {Our main aim is to study the influence of the initial conditions
     of a cloud in the intermediate/high-mass star formation process.}
   {We observed with the VLA, PdBI, and SMA the centimeter and
     millimeter continuum, \nth\,(1--0), and CO\,(2--1) emission
     associated with a dusty cloud harboring a nascent cluster with
     intermediate-mass protostars.}
   {At centimeter wavelengths we found a strong source, tracing a
     \uchii\ region, at the eastern edge of the dusty cloud, with a
     shell-like structure, and with the near-infrared counterpart
     falling in the center of the shell. This is presumably the most
     massive source of the forming cluster. About $15''$ to the west
     of the \uchii\ region and well embedded in the dusty cloud, we
     detected a strong millimeter source, MM1, associated with
     centimeter and near-infrared emission. MM1 seems to be driving a
     prominent high-velocity CO bipolar outflow elongated in the
     northeast-southwest direction, and is embedded in a ridge of
     dense gas traced by \nth, elongated roughly in the same direction
     as the outflow. We estimated that MM1 is an intermediate-mass
     source in the Class 0/I phase. 
     About $15''$ to the south of MM1, and still more deeply embedded in the
     dusty cloud, we detected a compact millimeter source, MM2, with
     neither centimeter nor near-infrared emission, but with water
     maser emission. MM2 is associated with a clump of \nth, whose
     kinematics reveal a clear velocity gradient 
     and additionally we found signposts of infall motions.  
     MM2, being deeply embedded within the dusty cloud, with an associated water
     maser but no hints of CO outflow emission, is
     an intriguing object, presumably of intermediate mass.}
   {The \uchii\ region is found at the border of a dusty cloud which
     is currently undergoing active star formation. Two
     intermediate-mass protostars in the dusty cloud seem to have
     formed after the \uchii\ region and have different properties
     related to the outflow phenomenon. Thus, a single cloud with
     similar dust emission and similar dense gas column densities
     seems to be forming objects with different properties, suggesting
     that the initial conditions in the cloud are not determining all
     the star formation process.}


   \maketitle
%

\begin{table*}
\caption{Main parameters of the VLA, PdBI, and SMA observations}
\centering
\footnotesize
\begin{tabular}{c c c c c c c c c c}
\hline\hline\noalign{\smallskip}
$\lambda$
&
&
&Epoch of
&Phase Calibrators
&Flux
&Beam
&P.A.
&cont. rms
&
\\
(cm)
&Obs
&Config
&Observation
&(Bootstrapped Flux, Jy)
&Calibrators
&($\arcsec\times\arcsec$)
&($\degr$)
&(m\jpb)
&Ref.
\\
\hline\hline\noalign{\smallskip}

6.0\phn	&VLA	&C\phe		&2006 Oct 27		&0102+584 ($2.897\pm0.011$)	&3C48		&$\phn5.2\times4.0$	&\phn$-4$	&0.03		&1	\\
\hline\noalign{\smallskip}
3.6\phn	&VLA	&B\phe		&1992 Jan 6		&2230+697 ($0.415\pm0.010$)	&3C48		&$\phn1.0\times0.8$	&$-66$		&0.10		&3	\\
3.6\phn	&VLA	&C\phe		&2006 Oct 27		&0102+584 ($3.172\pm0.029$)	&3C48		&$\phn3.1\times2.7$	&\phn$+3$	&0.02		&1	\\
3.6\phn	&VLA	&D\phe		&2004 Jul 10+13		&0102+584 ($1.885\pm0.003$)	&3C48		&$14.9\times8.5$	&$+76$		&0.04		&2	\\
\hline\noalign{\smallskip}
1.3\phn	&VLA	&C$^\mathrm{a}$	&1992 May 25		&0228+673 ($1.772\pm0.056$)	&3C48		&$\phn1.3\times1.0$	&$+50$		&0.6\phn	&3	\\
1.3\phn	&VLA	&D\phe		&2004 Jul 10+13		&0102+584 ($2.256\pm0.015$)	&3C48		&$\phn5.9\times4.5$	&$-21$		&0.07		&2	\\
\hline\noalign{\smallskip}
0.7\phn	&VLA	&D\phe		&2007 May 12		&0102+584 ($3.084\pm0.121$)	&3C286		&$\phn4.0\times3.0$	&$+64$		&0.17		&1	\\
\hline\noalign{\smallskip}
0.32	&PdBI	&CD\phe		&2004 Oct 17 + Dec 7	&0212+735 ($1.150$)		&MWC349		&$\phn4.3\times3.4$	&$+61$		&0.18		&1	\\
	&	&		&			&				&2145+067	&			&		&		&	\\
\hline\noalign{\smallskip}
0.12	&PdBI	&CD\phe		&2004 Oct 17 + Dec 7	&0212+735 ($0.620$)		&MWC349		&$\phn1.5\times1.2$	&$+58$		&1.0\phn	&1	\\
	&	&		&			&				&2145+067	&			&		&		&	\\
0.13	&SMA	&compact\phe	&2007 Jun 28		&0102+584 ($2.584$)		&Uranus		&$\phn3.2\times2.0$	&$+45$		&0.8\phn	&1	\\
	&	&		&			&0014+612 ($1.163$)		&		&			&		&		&	\\
\hline
\end{tabular}
\begin{list}{}{}
\item[$^\mathrm{a}$] Data from the VLA archive. The low sensitivity of
  the archival dataset does not allow to obtain better maps (when we
  combined the C and D configuration data) than previous VLA-D data
  maps (S\'anchez-Monge \et\ \cite{sanchezmonge2008}).
\item[References:] 
(1) This work, 
(2) S\'anchez-Monge \et\ \cite{sanchezmonge2008},
(3) VLA archival data.
\end{list}
\label{tobs}
\end{table*}

\section{Introduction \label{sint}}

It is well established that intermediate (2--8~\mo) and high-mass
  ($\gtrsim8$~\mo) stars form in cluster environments (\eg\ Kurtz
  \et\ \cite{kurtz2000}; Evans \et\ \cite{evans2009}),
and that at the very first stages of their formation, they are
embedded within dense and cold gas and dust from their original natal
cloud. However, it is not clear to what extent the initial conditions
of the natal cloud are determining the star formation story of the
cloud and the properties of the nascent stars of low,
  intermediate, and high mass, forming within it.
Regarding the star formation story, there is increasing evidence that
a single cloud can undergo different episodes of star formation, as
suggested by studies of deeply embedded clusters observed at spatial
scales similar to the cluster member separation ($\sim5000$~AU). In
these studies, the intermediate/high-mass young stellar objects
(YSOs) in the forming cluster seem to be in different evolutionary
stages (judging from the peak of their spectral energy distributions:
Beuther \et\ \cite{beuther2007}; Palau \et\ \cite{palau2007a},b;
Leurini \et\ \cite{leurini2007}; Williams \et\ \cite{williams2009}).


Concerning the properties of the intermediate/high-mass YSOs forming
within the cloud, a few studies toward massive star-forming regions
have revealed that two objects formed in the same environment may have
very different properties in the ejection of matter, with one YSO
driving a highly collimated outflow nearby another YSO driving an almost
spherical mass ejection (\eg\ Torrelles \et\ \cite{torrelles2001},
\cite{torrelles2003}; Zapata \et\ \cite{zapata2008}), indicating that
the initial conditions in the cloud may not be the only agent
determining the formation and evolution of the intermediate/high-mass
members in a forming cluster. While most of these cases have been
found at very high spatial scales ($\sim1$~AU), a broad observational
base in other regions and at other spatial scales is required to
properly understand the role of initial conditions of the cloud in the
cluster formation process and ultimately in the formation of
  intermediate/high-mass stars.

In this paper we show a high angular resolution study of a deeply
embedded cluster whose intermediate-mass protostars show differences
not only in their spectral energy distributions, but also in the
ejection phenomena associated with the YSOs.
The region, IRAS\,00117+6412, was selected from the list of
Molinari \et\ (\cite{molinari1996}) in a search for deeply embedded
clusters which are luminous ($>1000$~\lo) and nearby (distance
$<3$~kpc).
The IRAS source has a bolometric luminosity of 1400~\lo\ and is
located at a distance of 1.8~kpc (Molinari
\et\ \cite{molinari1996}). The millimeter single-dish image reported
by S\'anchez-Monge \et\ (\cite{sanchezmonge2008}) shows strong
emission with some substructure, tracing a dusty cloud, and the
centimeter emission reveals an ultra-compact \hii\ (\uchii) region,
associated with the brightest 2MASS source of the field at the eastern
border of the dusty cloud. The dusty cloud is associated with an
embedded cluster reported by Kumar \et\ (\cite{kumar2006}), two H$_2$O
masers spots (Cesaroni \et\ \cite{cesaroni1988}; Wouterloot
\et\ \cite{wouterloot1993}), and CO\,(2--1) bipolar outflow emission
(Zhang \et\ \cite{zhang2005}; Kim \& Kurtz \cite{kimkurtz2006}). All
this is suggestive of the dusty cloud harboring a \uchii\ region and a
deeply embedded stellar cluster forming in its surroundings.

We conducted high-sensitivity radio interferometric observations
in order to study the different millimeter sources embedded in the
dusty cloud. To properly characterize the protostars, we also studied the
distribution of dense gas, outflow and ionized gas emission. In the
present paper we show the first results obtained from \nth\ dense gas
and CO outflow emission, focusing mainly on the intermediate/high-mass
content of the protocluster.





\section{Observations \label{sobs}}


\subsection{Very Large Array observations \label{svla}}

IRAS~00117+6412 was observed with the Very Large Array
(VLA\footnote{The Very Large Array (VLA) is operated by the National
  Radio Astronomy Observatory (NRAO), a facility of the National
  Science Foundation operated under cooperative agreement by
  Associated Universities, Inc.}) at 6 and 3.6~cm with 5--8 EVLA
antennae in the array. The phase center of these observations was
$\alpha(\mathrm J2000)=00^{\mathrm h}14^{\mathrm m}27\fs725$, and
$\delta(\mathrm J2000)=+64\degr28\arcmin46\farcs171$. The integration
time was about 45~minutes at both wavelengths. Absolute flux
calibration was achieved by observing 3C48, with an adopted flux
density of 5.48~Jy at 6~cm, and 3.15~Jy at 3.6~cm.
The data reduction followed the VLA standard guidelines for
calibration and imaging, using the NRAO package AIPS. Images were
performed using different weightings. The robust parameter of Briggs
(\cite{briggs1995}) was set equal to 1 and 3 (almost natural)
respectively at 6 and 3.6~cm.  The configuration of the array,
observing dates, phase calibrator and bootstrapped fluxes, synthesized
beams and the continuum rms noise achieved are listed in Table 1.

The observations at 7~mm (together with \nh\ and SiO molecular line
observations, Busquet \et\ in prep.) were carried out with 9 EVLA
antennae in the array. In order to minimize the effects of atmospheric
fluctuations, we used the technique of \emph{fast switching} (Carilli
\& Holdaway \cite{carilliholdaway1997}) between the source and the
phase calibrator over a cycle of 120 seconds, with 80 seconds spent on
the target and 40 seconds on the calibrator. The on-source integration
time was about 1.8~hours.  Absolute flux and phase calibrations were
achieved by observing 3C286 (1.45~Jy) and 0102+584 (see Table~1). Data
reduction was performed following the VLA guidelines for the calibration
of high frequency data, using the NRAO package AIPS. The image was
constructed using natural weighting and tapering the \emph{uv}-data at
50~k$\lambda$ to increase the signal-to-noise ratio.

Searching the archive of the VLA, we found B-array data at 3.6~cm and
C-array data at 1.3~cm observed in 1992 (project AC295). The data were
reduced with the standard AIPS procedures.

\begin{figure*}[ht!]
\begin{center}
\begin{tabular}[b]{c}
	\epsfig{file=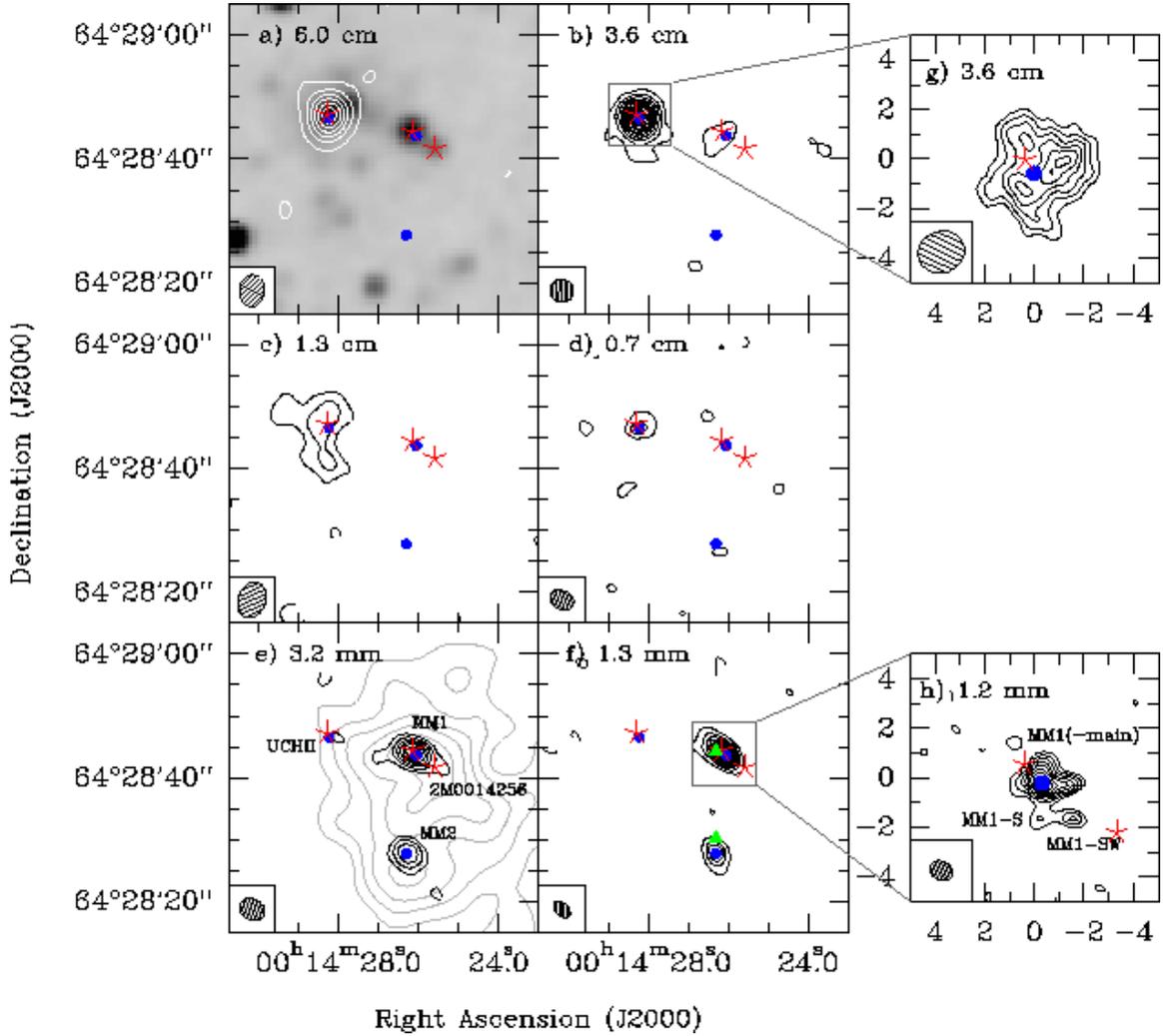, scale=0.95, angle=0} \\
\noalign{\bigskip}
\end{tabular}
\caption{IRAS~00117+6412 continuum maps.
\emph{(a)}: Grey-scale: 2MASS $K_\mathrm{s}$-band infrared
emission. Contours: VLA-C 6~cm continuum emission. Levels are $-3$,
and 3 to 33 in steps of 6, times 0.03~m\jpb.
\emph{(b)}: VLA-CD 3.6~cm continuum emission. Levels are $-3$,
and 3 to 30 in steps of 3, times 0.03~m\jpb.
\emph{(c)}: VLA-D 1.3~cm continuum emission. Levels are $-3$, 3, and 6
times 0.07~m\jpb\ (S\'anchez-Monge \et\ \cite{sanchezmonge2008}).
\emph{(d)}: VLA-D 0.7~cm continuum emission. Levels are $-3$,
3, and 6  times 0.17~m\jpb.
\emph{(e)}: Black contours: PdBI 3.2~mm continuum emission. Levels are
$-3$, and 3 to 21 in steps of 3, times 0.19~m\jpb. Grey contours:
IRAM\,30\,m continuum emission tracing the `dusty cloud', from S\'anchez-Monge
\et\ (\cite{sanchezmonge2008}). Levels are 4, 6, 8, 10, and 12 times the rms of the map,
8~m\jpb.
\emph{(f)}: SMA 1.3~mm continuum emission. Levels are $-3$, and 3 to
69 in steps of 6, times 0.62~m\jpb. Green triangles indicate the (approximate)
  position of the \water\ masers in the region (see Cesaroni \et\ \cite{cesaroni1988}).
\emph{(g)}: VLA-BCD 3.6~cm continuum emission. Levels are 3 to 8 in steps of 1, times 0.05~m\jpb.
\emph{(h)}: PdBI 1.2~mm continuum emission. Levels are 
$-3$, 3 to 11 in steps of 1, times 1.1~m\jpb.
In all panels, five-point red stars indicate the position of 2MASS
sources associated with the main centimeter and millimeter sources.
Blue dots indicate the position of the \uchii\ region, MM1-main, and
MM2, in order of descending declination.  Synthesized beams are
shown at the bottom-left corner of each panel, and correspond to those
indicated in Table~\ref{tmultiwave}.
}
\label{fi_cont}
\end{center}
\end{figure*}

\subsection{Plateau de Bure Interferometer observations \label{spdb}}

The continuum emission at 3 and 1.2~mm was observed simultaneously
together with the \nth\,(1--0), and CS\,(5--4) molecular transitions
with the Plateau de Bure Interferometer (PdBI\footnote{The Plateau de
  Bure Interferometer (PdBI) is operated by the Institut de
  Radioastronomie Millimetrique (IRAM), which is supported by
  INSU/CNRS (France), MPG (Germany), and IGN (Spain).}). The
observations were carried out during 2004 October 17 and December 7,
with the array in the D (5 antennae) and C (6 antennae)
configurations, respectively.  The phase center was $\alpha(\mathrm
J2000)=00^{\mathrm h}14^{\mathrm m}25\fs8$, and $\delta(\mathrm
J2000)=+64\degr28\arcmin43\farcs00$, and the projected baselines
ranged from 24.0 to 229.0~m.  The system temperatures were $\sim\!300$~K
for the receiver at 3.2~mm, and $\sim\!2000$~K for the receiver at
1.2~mm for both days. Atmospheric phase correction based on the 1.2~mm
total power was applied. The receiver at 3.2~mm was tuned to
93.17378~GHz (lower sideband) to cover the \nth\,(1--0) transition,
for which we used a correlator unit of 20~MHz of bandwidth and 512
spectral channels, providing a spectral resolution of 0.04~MHz
(0.13~\kms). The receiver at 1.2~mm was tuned to 244.93556~GHz (upper
sideband), covering the CS\,(5--4) transition, and the
  \chtoh\ 5$_0$--4$_0$ A+ and 5$_{-1}$--4$_{-1}$ E transitions. The CS
  and \chtoh\ line emission at 1.2~mm will be presented in a
  subsequent paper (Busquet \et, in prep.).
For the continuum measurements we used two
correlator units of 320~MHz in each band for both receivers. The FWHM
of the primary beam was $\sim\!54''$ at 3.2~mm, and $\sim\!22''$ at
1.2~mm.

\begin{table*}
\caption{Multiwavelength results for the intermediate-mass YSOs in the star-forming region IRAS~00117+6412}
\centering
\begin{tabular}{c l c c c c c c c}
\hline\hline\noalign{\smallskip}	
Wavelength
&
&Beam
&P.A.
&rms
&$I_\nu^\mathrm{peak}$$^\mathrm{a}$
&$S_\nu$$^\mathrm{a}$
&Deconv. Size
&P.A.
\\
(cm)
&Instrument
&($\arcsec\times\arcsec$)
&($\degr$)
&($\mu$Jy~beam$^{-1}$)
&(mJy~beam$^{-1}$)
&(mJy)
&($\arcsec\times\arcsec$)
&($\degr$)
\\
\hline\hline
\noalign{\smallskip}
\multicolumn{1}{l}{\uchii\ region}	&\multicolumn{2}{l}{$\alpha(J2000.0)=00^\mathrm{h}14^\mathrm{m}28\fs23$$^\mathrm{b}$}	\\
					&\multicolumn{2}{l}{$\delta(J2000.0)=+64\degr28\arcmin46\farcs6$$^\mathrm{b}$}		\\
\hline
\noalign{\smallskip}
6.0\phn\phe		&VLA-C		&$5.2\times4.0$		&\phn$-4$	&\phnn$30$
			&$1.21\pm0.03$	&$\phn1.72\pm0.08$	&$3.1\pm0.5\times2.7\pm0.4$	&\phn$95\pm50$	\\
3.6\phn\phe		&VLA-CD		&$3.8\times3.7$		&$+45$		&\phnn$30$
			&$1.03\pm0.03$	&$\phn1.88\pm0.09$	&$3.7\pm0.4\times3.1\pm0.4$	&\phn$10\pm60$	\\
3.6\phn\phe		&VLA-BCD	&$1.9\times1.7$		&$-70$		&\phnn$50$
			&$0.43\pm0.05$	&$\phn1.62\pm0.16$	&$3.9\pm0.4\times3.7\pm0.4$	&$180\pm60$	\\
1.3$^\mathrm{c}$\phn     &VLA-D		&$5.9\times4.5$		&$-21$		&\phnn$70$
			&$0.59\pm0.07$	&$\phn1.7\pm0.2$	&$5.9\pm1.8\times5\pm2$\phn\phn\phn    	&$120\pm50$ \\
0.7\phn\phe		&VLA-D		&$4.0\times3.0$		&$+64$		&\phn$170$
			&$1.35\pm0.17$	&\phn$1.4\pm0.3$	&$2.7\pm1.2\times0.0$\phn\phn\phn\phn\phe &$150\pm30$	\\
0.32\phe		&PdBI-CD	&$4.3\times3.4$		&$+61$		&\phn$180$
			&\ldots			&\phn$<1$		&\ldots				&\ldots		\\
0.13\phe		&SMA-comp	&$3.2\times2.0$		&$+45$		&\phn$600$
			&\ldots			&\phn$<4$		&\ldots				&\ldots		\\
0.12$^\mathrm{d}$	&PdBI-CD	&$1.5\times1.2$		&$+58$		&$1000$
			&\ldots			&$<25$		&\ldots				&\ldots		\\
\hline\hline
\noalign{\smallskip}
\multicolumn{1}{l}{MM1}	                &\multicolumn{2}{l}{$\alpha(J2000.0)=00^\mathrm{h}14^\mathrm{m}26\fs05$$^\mathrm{b}$}	\\
					&\multicolumn{2}{l}{$\delta(J2000.0)=+64\degr28\arcmin43\farcs7$$^\mathrm{b}$}		\\
\hline
\noalign{\smallskip}
6.0\phn\phe		&VLA-C		&$5.2\times4.0$		&\phn$-4$	&\phnn$30$
			&\ldots			&$<0.12$		&\ldots				&\ldots		\\
3.6\phn\phe		&VLA-CD		&$3.8\times3.7$		&$+45$		&\phnn$30$
			&$0.14\pm0.03$	&$0.17\pm0.07$	&$6.4\pm3.0\times3\pm2$\phn\phnn	&$120\pm30$	\\
3.6\phn\phe		&VLA-BCD	&$1.9\times1.7$		&$-70$		&\phnn$50$
			&$0.21\pm0.05$	&$0.15\pm0.06$	&$4.0\pm1.5\times2.2\pm1.5$	&$160\pm50$	\\
1.3\phn\phe		&VLA-D		&$5.9\times4.5$		&$-21$		&\phnn$70$
			&\ldots			&$<0.27$		&\ldots				&\ldots		\\
0.7\phn\phe		&VLA-D		&$4.0\times3.0$		&$+64$		&\phn$170$
			&\ldots			&$<0.69$		&\ldots				&\ldots		\\
0.32\phe		&PdBI-CD	&$4.3\times3.4$		&$+61$		&\phn$180$	
			&$4.16\pm0.19$	&$5.1\pm0.9$	&$3.2\pm0.4\times0.0$\phn\phn\phn\phn\phe       &\phn$75\pm10$	\\
0.13\phe		&SMA-comp	&$3.2\times2.0$		&$+45$		&\phn$600$
			&$51\pm1$\phn	&$71\pm14$	&$2.4\pm0.3\times1.0\pm0.2$	&\phn$50\pm10$	\\
0.12$^\mathrm{d}$	&PdBI-CD	&$1.5\times1.2$		&$+58$		&$1000$
			&$21\pm1$\phn	&$65\pm10$	&$2.9\pm0.4\times1.6\pm0.4$	&\phn$25\pm10$	\\
\hline\hline
\noalign{\smallskip}
\multicolumn{1}{l}{MM2}			&\multicolumn{2}{l}{$\alpha(J2000.0)=00^\mathrm{h}14^\mathrm{m}26\fs31$$^\mathrm{b}$}	\\
					&\multicolumn{2}{l}{$\delta(J2000.0)=+64\degr28\arcmin27\farcs8$$^\mathrm{b}$}		\\
\hline
\noalign{\smallskip}
6.0\phn\phe		&VLA-C		&$5.2\times4.0$		&\phn$-4$	&\phnn$30$
			&\ldots			&$<0.12$		&\ldots				&\ldots		\\
3.6\phn\phe		&VLA-CD		&$3.8\times3.7$		&$+45$		&\phnn$30$
			&\ldots			&$<0.12$		&\ldots				&\ldots		\\
3.6\phn\phe		&VLA-BCD	&$1.9\times1.7$		&$-70$		&\phnn$50$
			&\ldots			&$<0.21$		&\ldots				&\ldots		\\
1.3\phn\phe		&VLA-D		&$5.9\times4.5$		&$-21$		&\phnn$70$
			&\ldots			&$<0.27$		&\ldots				&\ldots		\\
0.7\phn\phe		&VLA-D		&$4.0\times3.0$		&$+64$		&\phn$170$
			&\ldots			&$<0.69$		&\ldots				&\ldots		\\
0.32\phe		&PdBI-CD	&$4.3\times3.4$		&$+61$		&\phn$180$	
			&$2.8\pm0.2$	&$3.2\pm0.6$	&$1.6\pm1.0\times1.0\pm1.5$	&$160\pm50$	\\
0.13\phe		&SMA-comp	&$3.2\times2.0$		&$+45$		&\phn$600$
			&$19\pm1$\phn	&$24\pm5$\phn	&$2.0\pm0.5\times0.0$\phn\phn\phn\phn\phe           &$170\pm20$	\\
0.12$^\mathrm{d}$	&PdBI-CD	&$1.5\times1.2$		&$+58$		&$1000$
			&$20\pm5$\phn	&$50\pm13$	&$3.5\pm1.0\times1.1\pm0.8$	&$140\pm30$	\\
\hline
\end{tabular}
\begin{list}{}{}
\item[$^\mathrm{a}$] Primary beam corrected (mainly affecting the
  millimeter data). In case of non-detection at one of the
  frequencies, an upper limit of 4$\sigma$ was used. Error in
  intensity is $1\sigma$. Error in flux density has been calculated as
  $\sqrt{(\sigma\sqrt{\theta_\mathrm{source}/\theta_\mathrm{beam}})^{2}+(\sigma_\mathrm{flux-scale})^{2}}$,
  where $\sigma$ is the rms noise level of the map,
  $\theta_\mathrm{source}$ and $\theta_\mathrm{beam}$ are the size of the
  source and the beam respectively, and $\sigma_\mathrm{flux-scale}$ is
  the error in the flux scale, which takes into account the uncertainty on
  the calibration applied to the flux density of the source
  ($S_\nu\times\%_\mathrm{uncertainty}$).
\item[$^\mathrm{b}$] Coordinates for the \uchii\ region, MM1, and MM2
  were measured in the VLA 3.6~cm, PdBI 1.2~mm, and SMA 1.3~mm images
  (Fig.~\ref{fi_cont}), respectively. Note that the coordinates of MM1
  correspond to the coordinates of MM1-main, listed in
  Table~\ref{tpdb1mm}.  
\item[$^\mathrm{c}$] Data from S\'anchez-Monge
  \et\ (\cite{sanchezmonge2008}). Deconvolved size and P.A. correspond
  to the component VLA~3 from S\'anchez-Monge
  \et\ (\cite{sanchezmonge2008}), which is the component associated
  with the easternmost 2MASS source shown in Fig.~\ref{fi_cont}c.
\item[$^\mathrm{d}$] 1.2~mm PdBI continuum map done with natural
  weighting; for flux densities and coordinates for the different
  subcondensations detected in the uniform-weighted map, see
  Table~\ref{tpdb1mm}.
\end{list}
\label{tmultiwave}
\end{table*}



Bandpass calibration was performed by observing 3C454.3 for October 17
and 2145+067 for December 7. We used the source 0212+735 to calibrate
the phases and amplitudes of the antennae for both days. The rms noise
of the phases was $< 50\degr$ for the data at 3.2~mm and $< 60\degr$ at
1.2~mm. The absolute flux density scale, calibrated with MWC349 on October 17
and 2145+067 on December 7, has an estimated uncertainty of
around 15\%. Data were calibrated using CLIC and imaged with MAPPING,
both part of the GILDAS software package.  Imaging of the 3 and 1.2~mm
emission was performed using natural and uniform weighting,
respectively. See Table~1 for details on the synthesized beams and the
rms noises.

\begin{table*}
\caption{Parameters of the 1.2~mm subcondensations associated with MM1 from the uniform-weighted PdBI map$^\mathrm{a}$}
\centering
\begin{tabular}{lccccccc}
\hline\hline\noalign{\smallskip}	
&\multicolumn{2}{c}{Position}
&$I_\mathrm{\nu}^\mathrm{peak}$$^\mathrm{b}$
&$S_\mathrm{\nu}$$^\mathrm{b}$
&Deconv. Size$^\mathrm{c}$
&P.A.$^\mathrm{c}$
&M$_\mathrm{env}$$^\mathrm{d}$
\\
\cline{2-3}
\noalign{\smallskip}
Source
&$\alpha (\rm J2000)$
&$\delta (\rm J2000)$
&(m\jpb)
&(mJy)
&($\arcsec\times\arcsec$)
&(\degr)
&(\mo)
\\
\hline\hline
\noalign{\smallskip}
MM1-main$^\mathrm{e}$	&00 14 26.051	&64 28 43.69	&$12.8\pm1.1$	 &$39\pm6$	  &$3.2\pm1.8\times0.3\pm0.3$	&$-89\pm\phn5$	&1.46	\\
			&		&		&		 &		  &$2.1\pm0.9\times1.0\pm0.3$	&\phn$+3\pm13$	&	\\
MM1-S			&00 14 26.065	&64 28 42.26	&\phn$4.7\pm1.1$ &\phn$5.4\pm1.4$ &$1.6\pm0.5\times0.8\pm0.2$	&$-30\pm30$	&0.20	\\
MM1-SW			&00 14 25.861	&64 28 42.25	&\phn$5.9\pm1.1$ &\phn$6.4\pm1.5$ &$1.3\pm0.4\times0.2\pm0.1$	&$+85\pm\phn9$	&0.24	\\
\hline
\end{tabular}
\begin{list}{}{}
\item[$^\mathrm{a}$] Synthesized beam of the uniform-weighted map is
  $0\farcs96\times0\farcs77$ with P.A.$=+66$\degr. The rms noise
  level of the map is 1.11~mJy~beam$^{-1}$.
\item[$^\mathrm{b}$] Primary beam corrected. Error in intensity and flux density as in Table~\ref{tmultiwave}.
\item[$^\mathrm{c}$] Deconvolved sizes and P.A. obtained from a fit
  with two Gaussian for MM1-main, and one single Gaussian for MM1-S and MM1-SW.
\item[$^\mathrm{d}$] Masses derived assuming a dust mass opacity
  coeffecient at 1.2~mm of 0.9~g\,cm$^{-1}$ (agglomerated grains with
  thin ice mantles in cores of densities $\sim10^6$~cm$^{-3}$,
  Ossenkopf \& Henning \cite{ossenkopfhenning1994}), and a dust temperature of 30~K.  The
  uncertainty in the masses due to the opacity law is estimated to be of
  a factor of 2.
\item[$^\mathrm{e}$] Throughout all the text, we refer to MM1-main as MM1, for simplicity.
\end{list}
\label{tpdb1mm}
\end{table*}

\subsection{Submillimeter Array observations \label{ssma}}

The Submillimeter Array (SMA\footnote{The SMA is a joint project
  between the Smithsonian Astrophysical Observatory and the Academia
  Sinica Institute of Astronomy and Astrophysics, and is funded by the
  Smithsonian Institution and the Academia Sinica.}; Ho \et\ \cite{ho2004}) in
the compact configuration was used to observe the 1.3~mm continuum
emission and the $^{12}$CO\,(2--1) molecular transition line (centered
at 230.538~GHz, upper sideband) on 2007 June 28. The phase center of
the observations was $\alpha\rm{(J2000)}=00^{\mathrm h}14^{\mathrm
  m}25\fs80$, and $\delta\rm{(J2000.0)}=+64\degr28\arcmin43\farcs0$,
and the projected baselines ranged from 9 to 78~k$\lambda$
(12--101~m).  System temperatures ranged between 80 and 200~K. The
zenith opacities, measured with the NRAO tipping radiometer located
at the Caltech Submillimeter Observatory, were good during the track,
with $\tau$(225~GHz)$\sim\!0.08$. The correlator, with a bandwidth of
1.968~GHz, was set to the standard mode, which provided a spectral
resolution of 0.8125~MHz (or 1.06~km~s$^{-1}$ per channel) across the
full bandwidth. The FWHM of the primary beam at 230~GHz was
$\sim\!56\arcsec$. The flagging and calibration of the data were done
with the MIR-IDL\footnote{The MIR cookbook by Charlie Qi can be found
  at http://cfa-www.harvard.edu/$\sim\!$cqi/mircook.html} package. The
passband response was obtained from observations of 3C454.3. The
baseline-based calibration of the amplitudes and phases was performed
using the sources 0102+584 and 0014+612. Flux calibration was set by
using Uranus, and the uncertainty in the absolute flux density scale
was $\sim\!20$\%. Imaging and data analysis were conducted using the
standard procedures in MIRIAD (Sault \et\ \cite{sault1995}) and AIPS (see Table~1
for details). The continuum was obtained by averaging all the
line-free channels of the upper sideband and the lower sideband. 




\section{Results \label{sres}}

\subsection{Centimeter continuum emission \label{srcm}}

We detected centimeter radio continuum emission at all wavelengths. In
Fig.~\ref{fi_cont}a we show the 6~cm continuum emission of the region.
The field is dominated by a strong and compact source associated with
the brightest infrared source in the field, which is the counterpart of
the \uchii\ region detected at 3.6~cm by S\'anchez-Monge
\et\ (\cite{sanchezmonge2008}). Additionally, the wide field of the VLA at 6~cm allows us
to detect two unresolved sources to the north and east of the
\uchii\ region. The northern one, with the coordinates of
$\alpha$(J2000.0)~=~00$^\mathrm{h}$14$^\mathrm{m}$03$^\mathrm{s}$.21,
and $\delta$(J2000.0)~=~+64$\degr$32$\arcmin$26$\arcsec$.5, has a primary beam
corrected flux density of $1.1\pm0.2$~mJy. The eastern one, with the coordinates of
$\alpha$(J2000.0)~=~00$^\mathrm{h}$14$^\mathrm{m}$57$^\mathrm{s}$.69,
and $\delta$(J2000.0)~=~+64$\degr$28$\arcmin$49$\arcsec$.6, has a primary beam
corrected flux density of $0.44\pm0.09$~mJy. Both sources are outside the
region shown in Fig.~\ref{fi_cont}a.

At 3.6~cm, we improved the angular resolution of previous observations
by a factor of five. These observations reveal two sources in the
field: an eastern source coincident with the \uchii\ region, and a
second and fainter source located $\sim\!15''$ to the
west. Fig.~\ref{fi_cont}b shows the resulting image after combining
the new VLA C-configuration observations with previous VLA
D-configuration observations (S\'anchez-Monge
\et\ \cite{sanchezmonge2008}). Since the combined dataset has a better
\emph{uv}-coverage, we recovered faint structure to the south of the
\uchii\ region similar to the double-source structure detected at
1.3~cm by S\'anchez-Monge \et\ (\cite{sanchezmonge2008}). The faint
source to the west of the \uchii\ region is coincident with the main
1.2~mm peak detected with the IRAM~30\,m telescope (S\'anchez-Monge
\et\ \cite{sanchezmonge2008}). By combining the 3.6~cm VLA-C and VLA-D
datasets with archival VLA-B data we improved the angular resolution up
to $\sim\!1.8''$.  With this angular resolution, the faint source
$15''$ to the west of the \uchii\ region is marginally detected, and
the \uchii\ region shows a shell morphology with three main peaks and
with the 2MASS source falling at the center of the peaks (see
Fig.~\ref{fi_cont}g).

In Fig.~\ref{fi_cont}c we show, for completeness' sake, the 1.3~cm
continuum emission map from S\'anchez-Monge \et\ (\cite{sanchezmonge2008}). At 7~mm we
detected one compact source at the position of the \uchii\ region
(Fig.~\ref{fi_cont}d).



In Table~\ref{tmultiwave} we summarize the main results of the sources
detected in the region at different wavelengths. The Table gives for
each source the coordinates, peak intensity, flux density,
deconvolved size and position angle.

\begin{figure*}[ht!]
\begin{center}
\begin{tabular}[b]{c}
	\epsfig{file=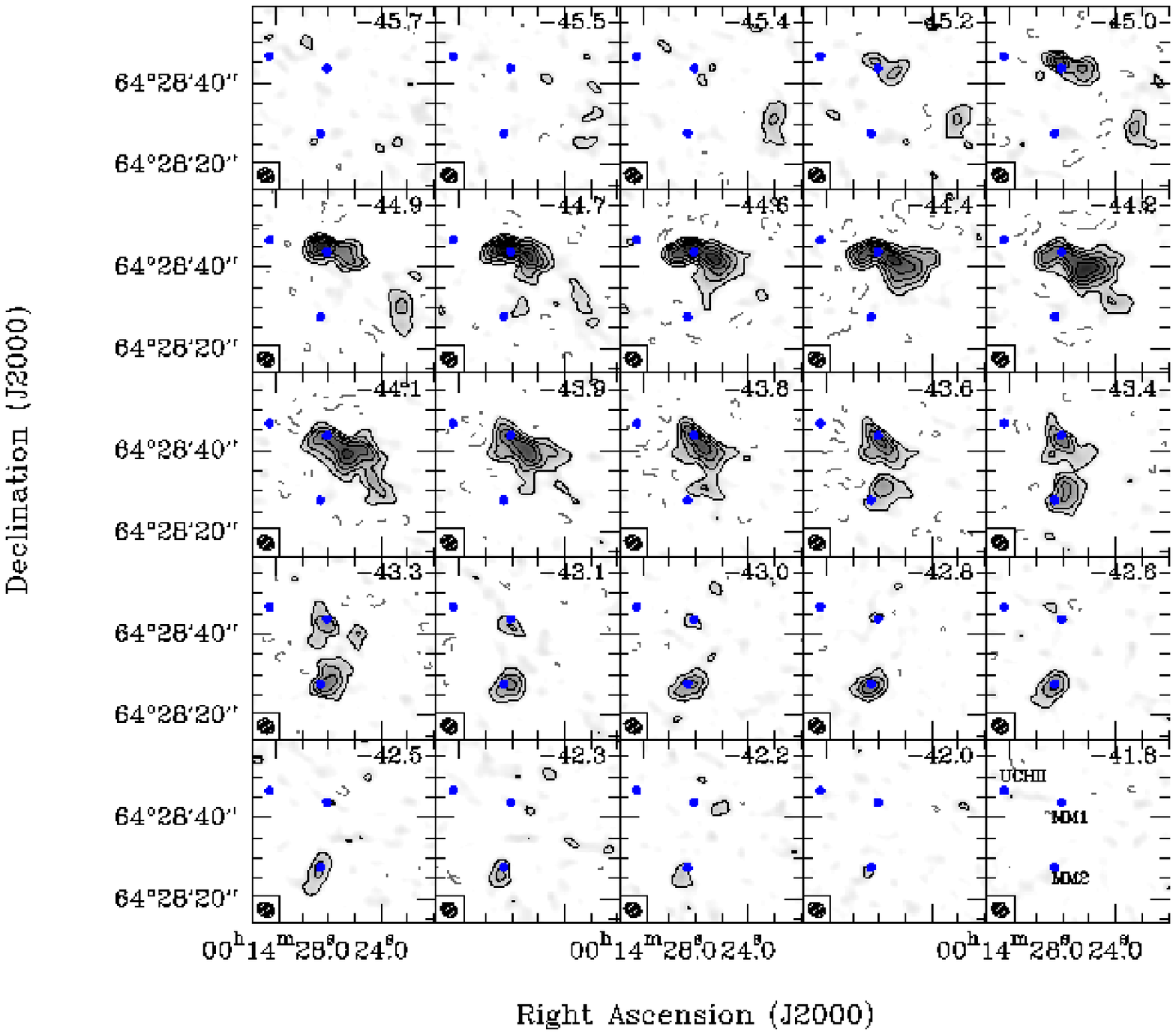, scale=1} \\
\noalign{\bigskip}
\end{tabular}
\caption{
Channel maps of the isolated \nth\,(1--0) hyperfine $F_1F=01\rightarrow12$ line, toward the
IRAS~00117+6412 region, with a spectral resolution of 0.16~\kms. The
central velocity of each channel is indicated in the upper right
corner, and the velocity to take as reference is $-44.3$~\kms. The
synthesized beam, shown in the bottom left corner of each panel, is
$4\farcs30\times3\farcs39$, at $\rm{P.A.}=60\fdg7$. Contours are $-6$,
$-3$, and 3 to 18 in steps of 3, times the rms noise, 0.013~Jy~beam$^{-1}$. Blue dots
indicate the position of the \uchii\ region, MM1, and MM2 as in
Fig.~\ref{fi_cont}.
}
\label{fn2hch}
\end{center}
\end{figure*}

\begin{figure}[ht!]
\begin{center}
\begin{tabular}[b]{c}
\hspace{-0.45cm}
       \epsfig{file=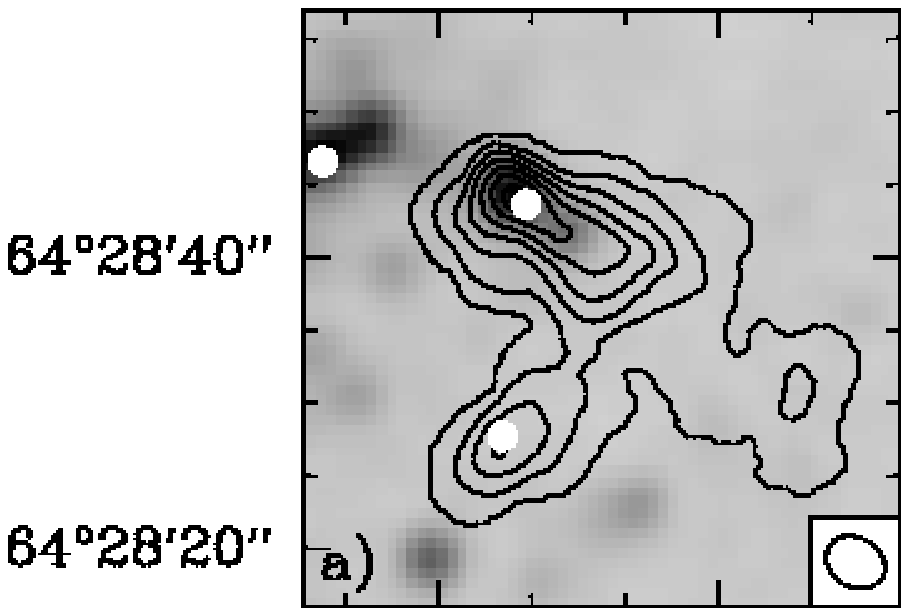, scale=0.7} \\
       \epsfig{file=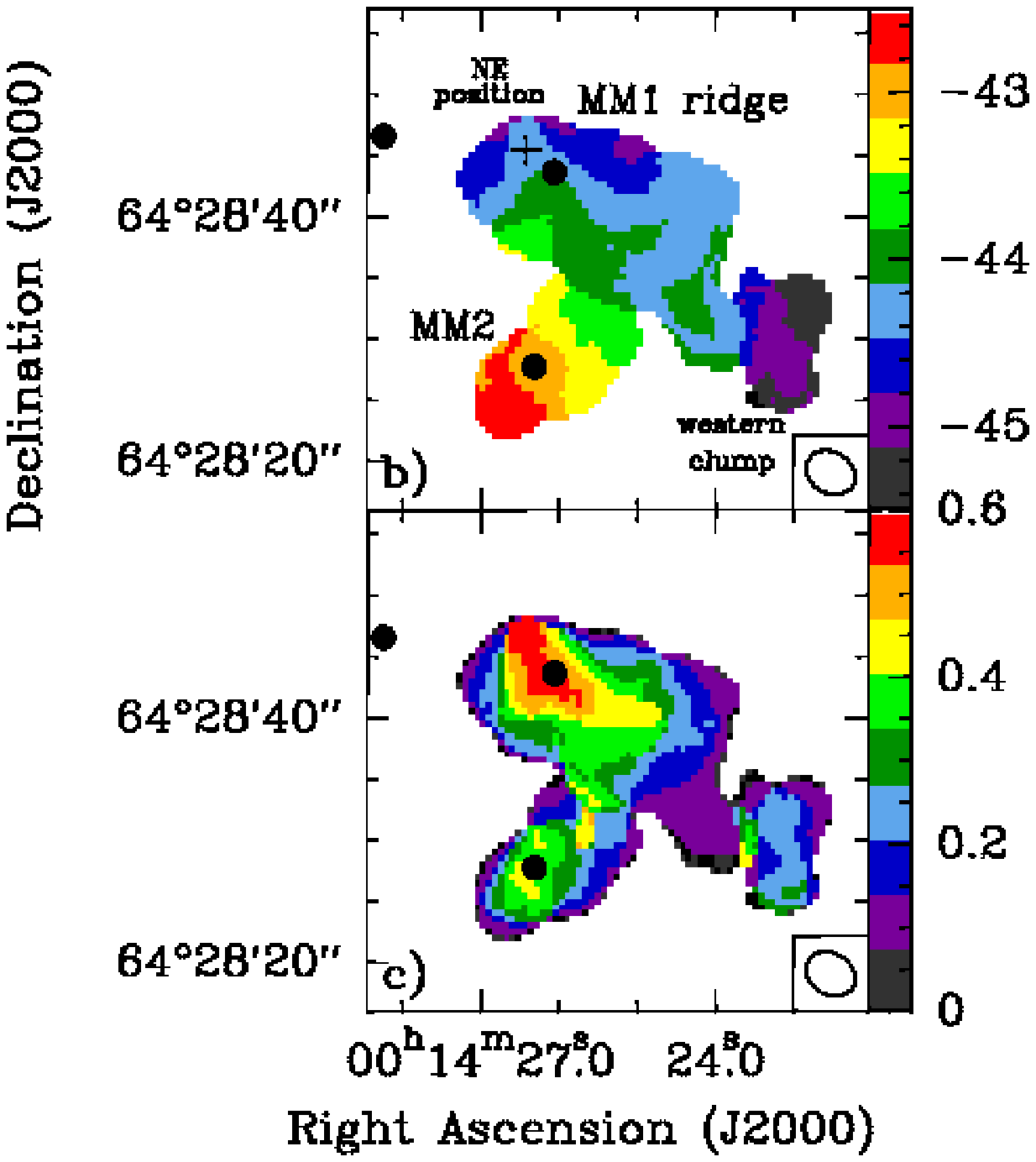, scale=0.7} \\
\noalign{\bigskip}
\end{tabular}
\caption{
 {\bf a)} Zero-order moment (integrated intensity) for the
 hyperfine $F_1F=01\rightarrow12$ line of \nth(1--0) toward
 IRAS~00117+6412. Contour levels range from 1 to 91\% of the peak
 intensity, 0.357~Jy~beam$^{-1}$~\kms, increasing in steps of
 15\%. Grey-scale: 2MASS $K_\mathrm{s}$-band infrared emission.
 {\bf b)} First-order moment (velocity) for the hyperfine
 $F_1F=01\rightarrow12$ line of \nth(1--0) (wedge units are \kms).
 Note that the velocity to take as reference is $-44.3$~\kms.
 Note also that the cross corresponds to the NE position of the MM1
 ridge ($4.0'', 2.5''$ from phase center), where a spectrum is shown
 in Fig.~\ref{fn2hspechfs} and is used to perform the pv-plot at
 PA=130\degr\ shown in Fig.~\ref{fn2hpv}a.  
 {\bf c)} Second-order moment (velocity dispersion) for the hyperfine
 $F_1F=01\rightarrow12$ of \nth(1--0) (wedge units are \kms). Note
 that the velocity dispersion must be multiplied by the factor
 $2\sqrt{2ln2}\simeq2.35$ to have linewidths.  In all panels, black/white
 dots indicate the position of the \uchii\ region, MM1, and MM2 as in
 Fig.~\ref{fi_cont}.  }

\label{fn2hm012}
\end{center}
\end{figure}

\subsection{Millimeter continuum emission \label{srmm}}

Figure~\ref{fi_cont}e shows the 3.2~mm PdBI continuum emission of the
region. There are two compact strong sources at 3.2~mm: the strongest
source, MM1, lying $\sim\!15''$ to the west of the \uchii\ region; and
the other source, MM2, being located $\sim\!15''$ to the south of MM1.
The peak of MM1 is associated with the faint 3.6~cm source shown in
Fig.~\ref{fi_cont}b and with a near-infrared 2MASS source. MM1 is
partially extended to the southwest, also spatially coincident with a
second infrared source detected in the 2MASS (J$00142558$+$6428416$),
hereafter 2M0014256 (cf.~Fig.~\ref{fi_cont}e). Note that the
$K_\mathrm{s}$-2MASS image shows no infrared emission toward
MM2 (cf.~Fig.\ref{fi_cont}a).

The SMA map at 1.3~mm shows two compact sources clearly associated
with MM1 and MM2 (Fig.~\ref{fi_cont}f). The 1.2~mm maps from the PdBI,
with an angular resolution three times better than the SMA images, but
only covering the region of MM1 within the primary beam of $22''$,
show that this source splits up into at least three subcondensations:
a compact core, MM1-main, elongated in the east-west direction and
with faint extensions towards the north and the south; a faint
4$\sigma$ source about $1''$ to the south of MM1-main, MM1-S; and a
faint source at 5$\sigma$ located $\sim2''$ to the southwest of
MM1-main, MM1-SW (see Fig.~\ref{fi_cont}h and
Table~\ref{tpdb1mm}).  We note that the PdBI 1~mm and SMA 1~mm
flux densities are completely consistent, if we take into account that
the PdBI is filtering out emission at smaller scales than the
SMA. Throughout all the paper we will refer to MM1-main as MM1.

In Table~\ref{tmultiwave} we summarize the main results of the sources
detected in the region at millimeter wavelengths (see Sect.~\ref{srcm}
for a description of the table). For both MM1 and MM2, we estimated
the mass of the dust component assuming dust temperatures of 30~K and
20~K for MM1 and MM2, respectively (as a first approximation, since
MM1 is associated with near-infrared emission, while MM2 is not), and
a dust mass opacity coefficient at 1.3~mm of 0.9~g\,cm$^{-1}$
(agglomerated grains with thin ice mantles in cores of densities
$\sim10^6$~cm$^{-3}$, Ossenkopf \& Henning
\cite{ossenkopfhenning1994}). With these assumptions, we estimated a
mass for MM1 of $\sim3.0$~\mo, and a mass for MM2 of
$\sim\!1.7$~\mo\ (both derived from the (SMA) flux at 1.3~mm given in
Table~\ref{tmultiwave}). In order to properly estimate the spectral
index between 3.2~mm (PdBI) and 1.3~mm (SMA), we made the SMA image
using the same $uv$-range as the PdBI data (9--72~k$\lambda$). By
measuring the flux densities in the maps in the common $uv$-range, we
obtained a spectral index of $3.1\pm0.2$ for MM1 and $2.7\pm0.2$ for
MM2. Finally, we estimated the fraction of missing flux resolved out
by the SMA at 1.3~mm by comparing the total flux of MM1 plus MM2 with
the flux density measured with the IRAM\,30\,m Telescope of 1.2 Jy
(S\'anchez-Monge \et\ \cite{sanchezmonge2008}), and found that the
fraction of flux filtered out by the SMA is 92\%. 




\subsection{N$_2$H$^+$ emission \label{srn2h}}

Figure~\ref{fn2hch} shows the \nth~(1--0) channel maps of the
hyperfine $F_1F=01\rightarrow12$ line (hereafter, the `isolated' hyperfine) toward
IRAS\,00117+6412. Close to the systemic velocity\footnote{The systemic
  velocity for IRAS\,00117+6412 is $-36.3$~\kms\ (Molinari \et\ \cite{molinari1996};
  Zhang \et\ \cite{zhang2005}), and the \nth\,(1--0) hyperfine adopted as
  reference line in this work was $F_1F=23\rightarrow12$. However, since the
  hyperfine adopted as reference is blended with two other hyperfines,
  we will make our analysis with the hyperfine which is isolated, the
  $F_1F=01\rightarrow12$ line, which is at 8~\kms\ with respect to the line
  adopted as reference. Therefore, the reference velocity
  (corresponding to the systemic velocity) in the analysis of the
  \nth\ emission from the isolated hyperfine is $-36.3-8.0=-44.3$~\kms.}
and spanning $\sim\!2$~\kms, the emission
appears as a ridge at the MM1 position (labeled as the MM1 ridge) and is
elongated roughly in the east-west direction (PA$\sim58^\circ$). While
at redshifted velocities the emission is only associated with MM2
(hereafter, the MM2 clump), the most blueshifted velocities show some
faint emission $\sim\!20''$ towards the west of MM2 (western
clump). The zero-order moment map integrated for the isolated hyperfine
of \nth(1--0) is presented in Fig.~\ref{fn2hm012}a,
overlaid on a $K_\mathrm{s}$-band image of 2MASS. The MM1 ridge has a
length of $20''$ and the crest is coinciding well with the peak and
elongation of the 3.2~mm continuum emission (Fig.~\ref{fi_cont}e). As
for the MM2 clump, it has a size of $\sim\!10''$, with the peak
position also coinciding with the millimeter peak, and is elongated in
the southeast-northwest direction (PA$\sim132^\circ$).  Finally, it is
worth noting that the MM2 clump and the western clump fall at a region
which is completely dark in the near-infrared, as seen in the
$K_\mathrm{s}$-2MASS image of Fig.~\ref{fn2hm012}a.

\subsection{CO\,(2--1) emission \label{srco}}

\begin{figure*}[ht!]
\begin{center}
\begin{tabular}[b]{c}
	\epsfig{file=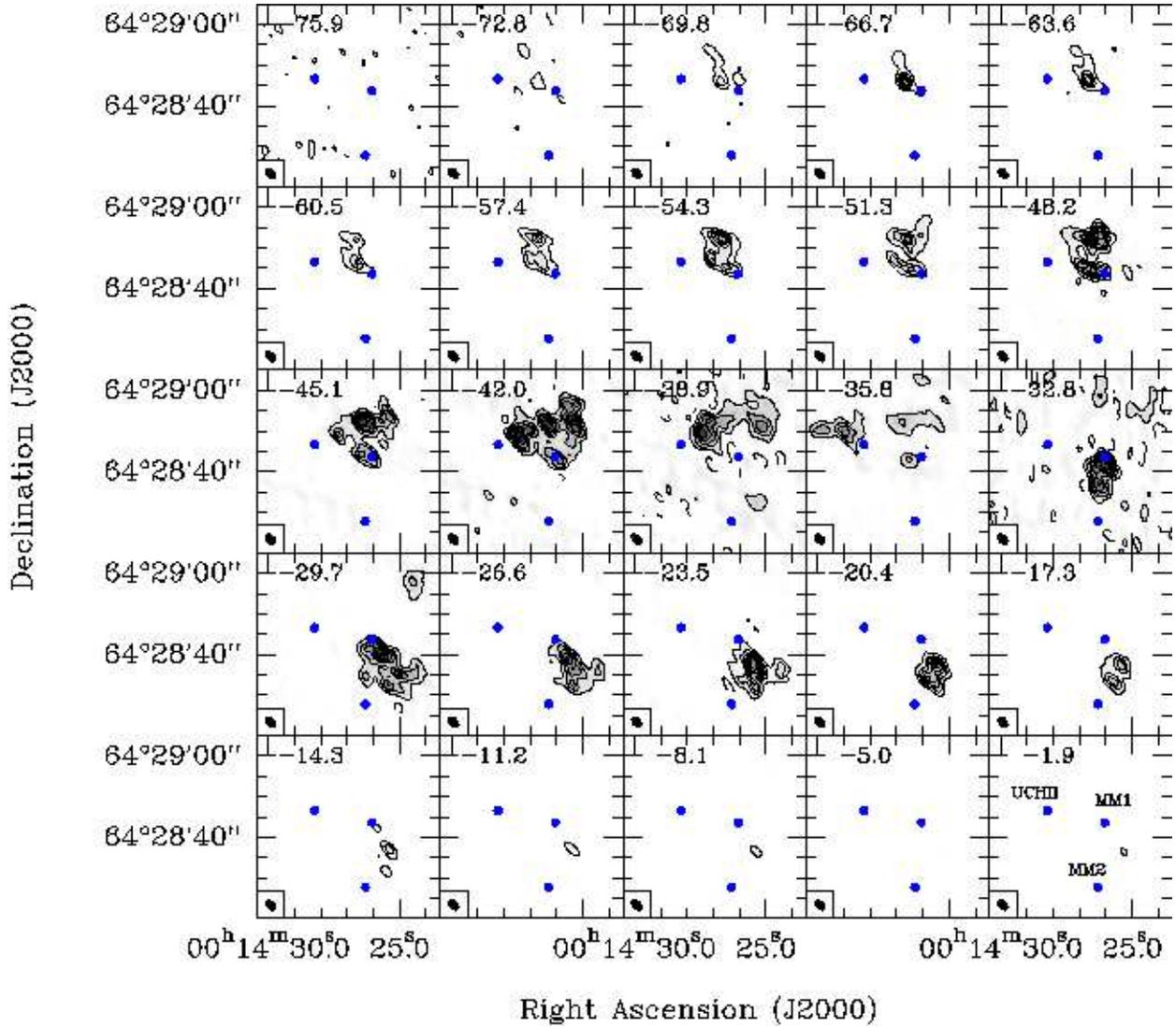, scale=1} \\
\noalign{\bigskip}
\end{tabular}
\caption{CO\,(2-1) channel maps of the IRAS~00117+6412 region, averaged over
3.1~\kms\ wide velocity intervals. The central velocity of each channel is
indicated in the upper left corner, and the systemic velocity is $-36.3$~\kms.
Symbols are the same as in Fig.~\ref{fi_cont}. The synthesized beam, shown
in the bottom left corner of each panel, is $2\farcs92\times1\farcs78$, at
$\rm{P.A.}=44\fdg7$. Contours are $-4$, and 4 to 40 in steps of 6, times the rms
noise, 0.1~\jpb.}
\label{fi_12co_channel}
\end{center}
\end{figure*}

Channel maps of the CO\,(2--1) emission are displayed in
Fig.~\ref{fi_12co_channel}. The CO\,(2--1) emission appears spanning a
wide range of velocities, from $-70$ up to $-14$~\kms\ (systemic
velocity at $-36.3$~\kms).  Blueshifted emission appears in different
clumps toward the north and northeast of MM1, while redshifted
emission appears toward the southwest of MM1, suggesting a
bipolar structure centered on MM1.

The spectrum of the emission integrated over all the region is shown
in Fig.~\ref{fi_12co_spectra_1}, together with a preliminary spectrum
of $^{13}$CO (Busquet \et, in prep.). The CO spectrum shows a dip 
from $-6$ up to $+6$~\kms\ with respect to the systemic velocity,
which could be due to self-absorption by cold foreground gas or
opacity effects, as the dip is coincident with the peak of the
$^{13}$CO line. We note that the dip could be partially produced by
the missing short-spacing information in the interferometer data as
well. In order to make a rough estimate of the fraction of flux
filtered out by the SMA (due to the lack of $uv$ sampling at spacings
smaller than 9~k$\lambda$), we compared the single-dish spectrum (from
Zhang \et\ \cite{zhang2005}) at certain velocities with the SMA
spectrum (extracted from the channel maps of
Fig.~\ref{fi_12co_channel} convolved to the single-dish beam, of
$30''$) at the position of the IRAS source.  The missing flux is
around 99\% for the systemic velocities, indicating that at these
velocities the CO emission mainly comes from structures much larger
than $\sim10''$, which is the largest angular scale observable by an
interferometer whose shortest baseline is $\sim9$~k$\lambda$ (see
Appendix for details). On the other hand, the missing flux
decreases as one moves to higher velocities. For example, the fraction
of missing flux goes down to 50\% at $-7$ and 5~\kms\ (with respect to
the systemic velocity) for the blueshifted and redshifted sides,
respectively; and the SMA recovers all the single-dish flux at $-13$
and 7~\kms\ with respect to the systemic velocity. This indicates that
the high-velocity CO\,(2--1) emission has a characteristic source size
$\lesssim10''$.

\section{Analysis \label{sa}}

\subsection{\nth\ kinematics: moments and pv-plots \label{san2hmom}}

In order to study the kinematics of the dense gas as traced by \nth,
we computed the first-order moment for the isolated hyperfine of
\nth(1--0), which does not suffer from blending with other hyperfines
(Fig.~\ref{fn2hm012}b). The first-order moment map shows a velocity
gradient in the MM1 ridge, spanning velocities of $-43.5$ to
$-45.0$~\kms, centered approximately on MM1, and perpendicular to the
elongation of the ridge. Another clear velocity gradient can be seen
toward MM2 at red velocities, from $-42$ to $-44$~\kms. Note that both
velocity gradients observed toward MM1 and MM2 have the blueshifted
velocities toward the northwest and the redshifted velocities toward
the southeast. In Fig.~\ref{fn2hm012}c we show the second-order moment
map of the region, tracing the velocity dispersion, which is largest
toward MM1 and MM2, and in general along the MM1 ridge, reaching
(linewidth) values of 0.9--1.4~\kms, much larger than the thermal
linewidth for \nth, of around 0.22~\kms\ at 30~K. The western clump is
the more quiescent emission in the region, with linewidths of up to
0.6~\kms. In addition, the first-order moment reveals an abrupt change
in velocity in the western clump (Fig.~\ref{fn2hm012}b), which could
be indicative of the clump being kinematically decoupled from the MM1
ridge and the MM2 clump.

\begin{figure}[ht]
\begin{center}
\begin{tabular}[b]{c}
\noalign{\bigskip}
	\epsfig{file=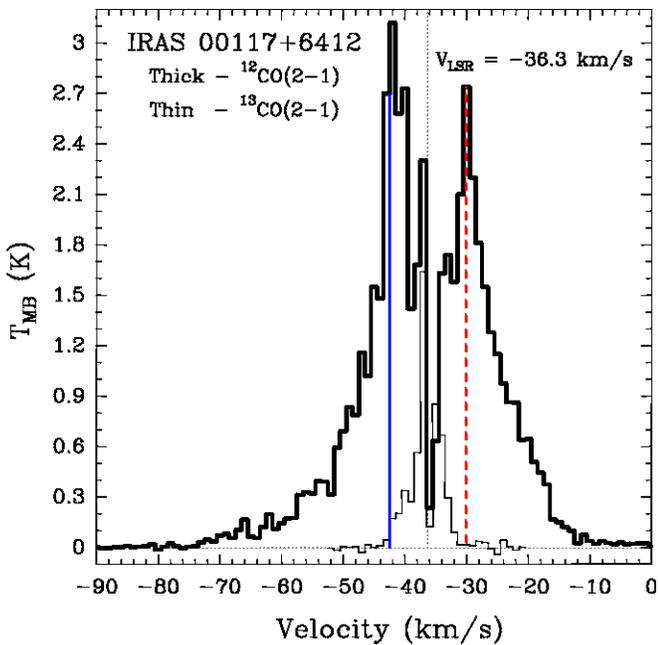, scale=0.5} \\
\noalign{\bigskip}
\end{tabular}
\caption{Spectrum of the $^{12}$CO\,(2-1) (thick line) and
  $^{13}$CO\,(2-1) (thin line) emission in the IRAS~00117+6412 region,
  averaging over all emission of outflow lobes. Blue (solid) and
  red (dashed) vertical lines indicate the range of velocities (red
  wing: $-30.0$~\kms\ up to $-8.0$~\kms, and blue wing:
  $-72.0$~\kms\ up to $-42.5$~\kms) used to estimate the parameters of
  the outflow.}
\label{fi_12co_spectra_1}
\end{center}
\end{figure}

\begin{figure}[ht!]
\begin{center}
\begin{tabular}[b]{c}
      \epsfig{file=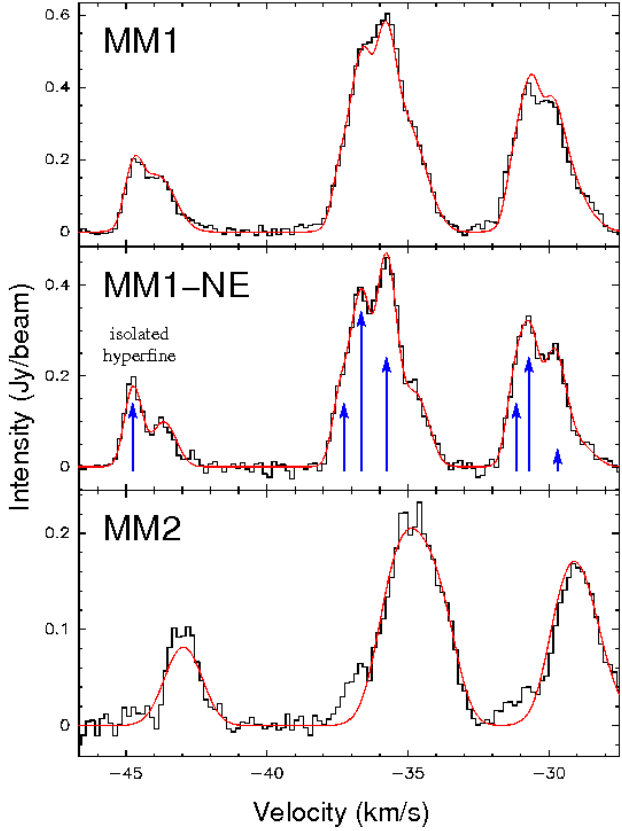, scale=0.6, angle=0}\\
\noalign{\bigskip}
\end{tabular}
\caption{
\emph{Top}: \nth\,(1--0) hyperfine spectra toward MM1.
\emph{Middle}: idem toward the NE position of the MM1 ridge ($4.0,2.5$
with respect to the phase center; see Fig.~\ref{fn2hm012}b). 
\emph{Bottom}: idem toward MM2.
For MM1, and the NE position, we fitted the hyperfine structure with
two velocity components, and for MM2 we used one single velocity
component.
In the middle panel, the blue arrows indicate the position of the seven
hyperfines (with statistical weights from Womack
\et\ \cite{womack1992} and frequencies from Caselli
\et\ \cite{caselli1995}) for the velocity component at
$-36.7$~\kms\ (Table~\ref{tn2hh}). The different hyperfine lines
correspond, from left (blue velocities) to right (red velocities), to
the quantum numbers: $F_1F=01\rightarrow12$ (isolated line),
$21\rightarrow11$, $23\rightarrow12$ (reference line),
$22\rightarrow11$, $11\rightarrow10$, $12\rightarrow12$, and
$10\rightarrow11$.
}
\label{fn2hspechfs}
\end{center}
\end{figure}

\begin{table*}[t]
\caption{Parameters of the hyperfine fits to the \nth\,(1--0) line}
\centering
\footnotesize
\begin{tabular}{lccccccc}
\hline\hline\noalign{\smallskip}
&&$A\tau_\mathrm{m}^\mathrm{b}$
&Velocity$^\mathrm{c}$
&Linewidth$^\mathrm{c}$
&&$\Tex$
&$N$(\nth)
\\
Object
&Line$^\mathrm{a}$
&(K\,\kms)
&(\kms)
&(\kms)
&$\tau_\mathrm{m}$$^\mathrm{b}$
&(K)
&(cm$^{-2}$)
\\
\noalign{\smallskip}
\hline\noalign{\smallskip}
MM1     &1	&$0.062\pm0.003$   &$-36.7$     &$0.66$   &$0.20\pm0.02$    &3.1  &$6.7\times10^{11}$\\
        &2      &$0.060\pm0.001$   &$-35.9$     &$1.40$   &$\lesssim0.1$    &3.4  &$8.5\times10^{11}$\\
MM1-NE  &1	&$0.072\pm0.003$   &$-36.7\pm0.1$&$0.68\pm0.02$&$0.15\pm0.02$&3.3  &$5.8\times10^{11}$\\
        &2      &$0.038\pm0.001$   &$-35.6\pm0.1$&$0.98\pm0.04$&$\lesssim0.1$&3.2  &$5.3\times10^{11}$\\
MM2     &       &$0.031\pm0.001$   &$-34.9\pm0.1$&$1.49\pm0.05$&$\lesssim0.1$ &3.1 &$7.6\times10^{11}$\\
\hline
\end{tabular}
\begin{list}{}{}
\item[$^\mathrm{a}$] For MM1 and the NE position of the MM1 ridge,
  the hyperfine structure is fitted with two velocity components.
  For MM2 the fit is made with one single velocity component.
\item[$^\mathrm{b}$]
  $A=f[J_\nu(T_\mathrm{ex})-J_\nu(T_\mathrm{bg})]$. $\tau_\mathrm{m}$
  is the optical depth of the main hyperfine, adopted to be the
  $F_1F=23\rightarrow12$ line. The otpical depth of the isolated line
  can be obtained by multiplying $\tau_\mathrm{m}$ by the factor 3/7.
\item[$^\mathrm{c}$] When errors are not given, the velocity and
  linewidth have been fixed from a fit of two Gaussian to the isolated
  line. When errors are given, the fit of two Gaussian to the isolated line
  was used to provide initial values for the velocity and linewidth
  of the hyperfine fit.
\end{list}
\label{tn2hh}
\end{table*}

In Fig.~\ref{fn2hspechfs} we show the \nth\ spectra for MM1, for a
position in the northeastern (NE) side of the MM1 ridge (see positions
in Fig.~\ref{fn2hm012}b), and for MM2. Note that the lines are broad,
as can be seen in the isolated hyperfine line (at $\sim -44$~\kms),
and suggestive of a double velocity component. See Sects.~\ref{samm1}
and \ref{samm2} for a further analysis and discussion on these
spectra.

\begin{figure}[ht]
\begin{center}
\begin{tabular}[b]{cc}
       \epsfig{file=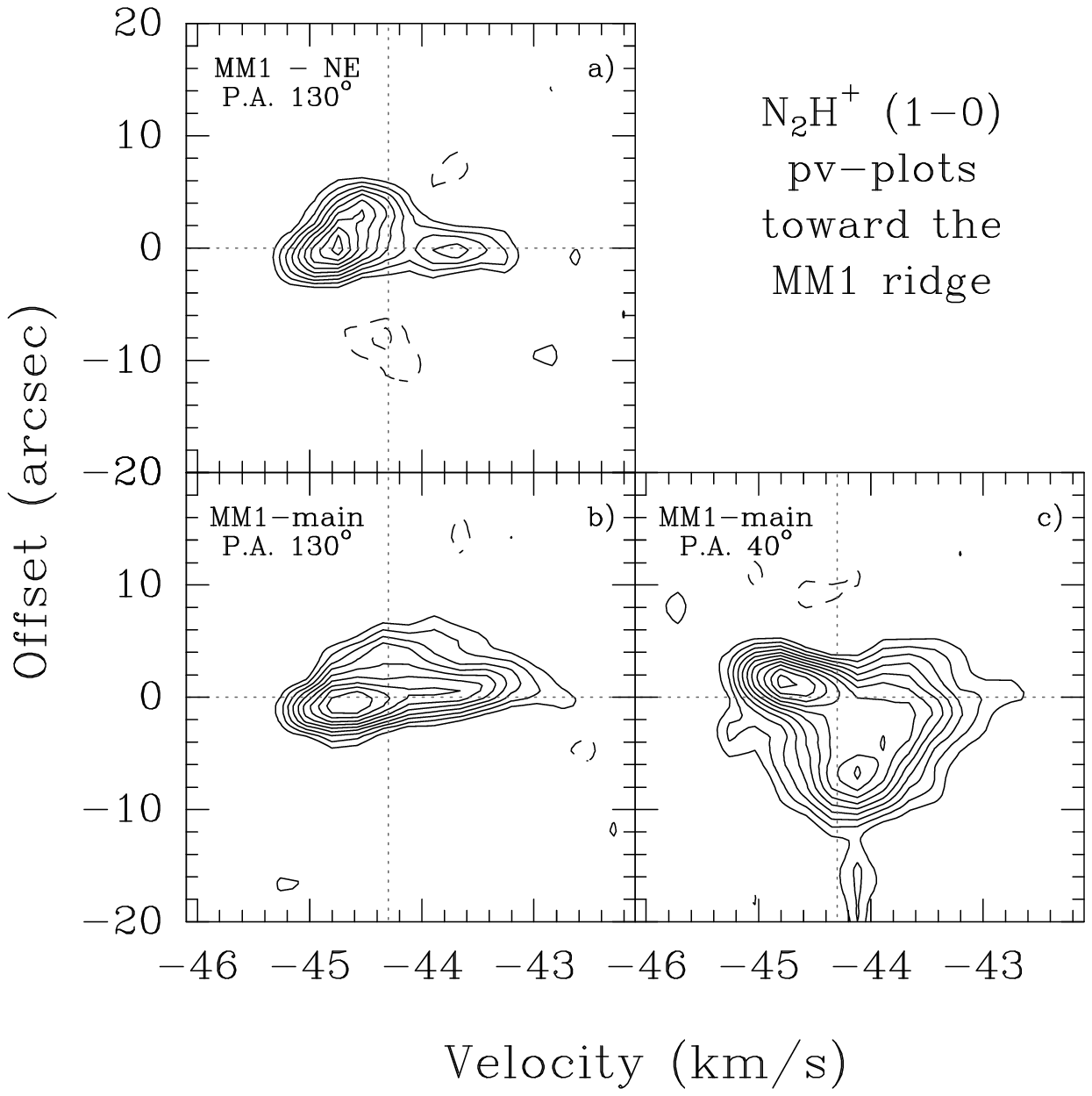, scale=0.65}\\
\noalign{\bigskip}
\end{tabular}
\caption{
 Position-velocity (pv) plots of the \nth\ isolated line between $-46$ and
 $-42$~\kms, toward positions of the MM1 ridge.
 {\bf a)} Pv-plot in a cut with PA=130\degr, and centered on the NE
 position of the MM1 ridge ($4\farcs0, 2\farcs5$ with respect to the
 phase center, see Fig.~\ref{fn2hm012}b).
 {\bf b)} Pv-plot in a cut with PA=130\degr, and centered on MM1.
 {\bf c)} Pv-plot in a cut with PA=40\degr, centered on MM1.
 Note that the velocity to take as reference is $-44.3$~\kms. For all panels,
 contours are $-3$, $-2$, 2 to 10 in steps of 1, times the rms noise level,
 0.02~\jpb.
}
\label{fn2hpv}
\end{center}
\end{figure}

\begin{figure}[h!]
\begin{center}
\begin{tabular}[b]{c}
        \epsfig{file=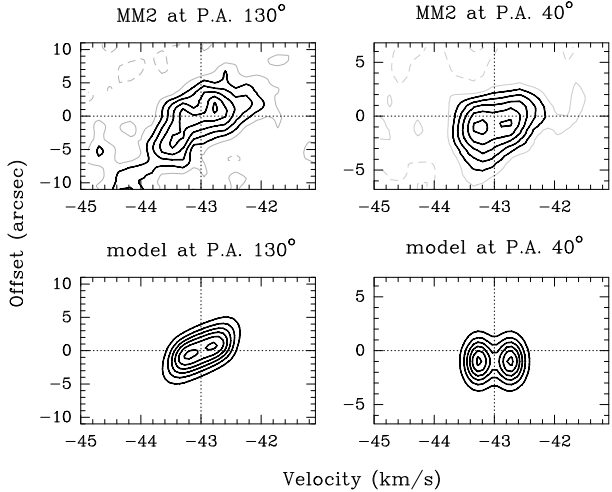, scale=0.45, angle=0}\\
\noalign{\bigskip}
\end{tabular}
\caption{
Position-velocity (pv) plot of the \nth\ isolated line toward MM2 in a cut of
PA=130\degr\ (top left) and PA=40\degr\ (top right) and the
corresponding modeled emission (bottom) of a disk structure with
infall and rotation, adopting the parameters shown in
Table~\ref{tn2hpvinfrot} for the case of an inner radius of $\sim1''$ (we
shifted the disk with respect to the MM2 position by $-1''$ in the
PA=40 axis). For the top panels, (grey) black contours are ($-16$\%,
16\%,) 32\%, 48\%, 64\%, 80\%, and 96\% of the peak emission,
0.146~\jpb\ for the top-left panel, and 0.148~\jpb\ for the top-right
panel. For the bottom panels, contours are 16\%, 32\%, 48\%, 64\%,
80\%, and 96\% of the peak emission, which is in arbitrary units.
}
\label{fn2hpvinfrot}
\end{center}
\end{figure}

We additionally built position-velocity (pv) cuts to further study the
kinematics of the region. In Fig.~\ref{fn2hpv} we show two cuts in
the direction of the velocity gradient seen in the MM1 ridge (at
PA$\sim130$\degr, for MM1 and a position to the NE of MM1, see
positions in Fig.~\ref{fn2hm012}b), and one cut along the elongation
of the ridge (PA=40\degr). The three cuts show emission spanning from
$\sim-43$ to $\sim-45$~\kms, with one compact component at the blueshifted
velocities of ($-44.2$)--($-45.0$)~\kms, and a broad red component at
$\sim-43.5$~\kms. 
The emission from the NE position and from MM1 follow the same general
trend of a blue component with a small velocity gradient and a red
broad component with no gradient (better seen in the cut of the NE
position, Fig.~\ref{fn2hpv}a). For the perpendicular cut (PA=40\degr,
centered on MM1, Fig.~\ref{fn2hpv}c), we also find a velocity
gradient with the redshifted velocities at negative values, or to the
southwest of the cut, and the blueshifted toward the northeast.
Regarding MM2, the velocity gradient is also measured at a
PA=130\degr, coincident with the elongation of the MM2 clump
(Fig.~\ref{fn2hm012}b), and is well seen in the pv-plot at
PA=130\degr, with a double peak
(Fig.~\ref{fn2hpvinfrot}-top). However, the perpendicular cut does not
show any gradient, although the double peak is also there.  We refer
the reader to the subsequent sections for the interpretation of the
kinematics of the \nth\ dense gas toward MM1 and MM2.


\subsection{\nth\ kinematics: MM1 ridge \label{samm1}}

In order to fit the hyperfine spectra toward MM1 and the NE position
of the ridge (spectra shown in Fig.~\ref{fn2hspechfs}), we assumed two
velocity components. We initially fitted two Gaussian to the isolated
line to estimate the initial values of the velocities and linewidths
for the two velocity components.  With these initial values, we fitted
the hyperfine spectra with two velocity components for the NE
position and MM1. The results are listed in Table~\ref{tn2hh} and
shown in Fig.~\ref{fn2hspechfs}. The fits can reproduce the data
reasonably well. 

For MM1, the derived two velocity components at $-36.7$~\kms, with a
linewidth of $\sim0.7$~\kms, and $-35.9$~\kms, with a linewidth of
$\sim1.4$~\kms, were already seen in the pv-plots of MM1 at
PA=130\degr\ shown in Fig.~\ref{fn2hpv} (note that in
Fig.~\ref{fn2hpv} we show the isolated line, which was \emph{not}
taken as the reference line, and the component fitted at
$-36.7$~\kms\ corresponds to $-44.7$~\kms, while the component at
$-35.9$~\kms\ corresponds to $-43.9$~\kms\ in the pv-plot). For the NE
position of the ridge, the two velocity components are found again
around $-36.7$ and $-35.6$~\kms, and the redshifted component has a
broader linewidth than the blueshifted component, as in the MM1
spectrum. We note that the red component is optically thinner than the
main component. This red component, which shows no velocity gradient,
could be tracing a filament intercepting the line of sight maybe
connecting the MM1 ridge and MM2, as the velocity of this red
component is similar to the velocity of MM2. Finally, we showed in
Sect.~\ref{san2hmom} (Fig.~\ref{fn2hpv}a) that the blue component
(corresponding to the strongest component in the spectra) has a small
velocity gradient which could be tracing the global rotation of the
MM1 ridge.

In Table~\ref{tn2hh} we also list the excitation temperature derived
from the fits and the \nth\ column density calculated assuming that
the filling factor is $\sim1$ and following Caselli
\et\ (\cite{caselli2002b}). We note that the excitation temperature of
$\sim3$~K, and the \nth\ column densities are similar to the values
found in low-mass star-forming regions (\eg\ Caselli
\et\ \cite{caselli2002a}; Chen \et\ \cite{chen2007}; Kirk
\et\ \cite{kirk2009}), and slightly smaller than the values found in
massive star-forming regions observed with interferometers (\eg\ Palau
\et\ \cite{palau2007a}; Beuther \& Henning
\cite{beutherhenning2009}). The small excitation temperature obtained
is suggestive of either low density and/or cold gas.
Note however that our values could be affected by the missing flux
problem caused by the lack of short $uv$-spacings in the
interferometric data (for the PdB, the largest angular scale
detectable is $\sim11''$, see Appendix, while in Palau
\et\ (\cite{palau2007a}) and Beuther \& Henning
(\cite{beutherhenning2009}), the largest angular scale was 20--$30''$,
which means that our PdB data are more affected by the missing flux
problem).


\begin{table*}
\caption{Output parameters of the position-velocity plot modeled emission for MM2}
\centering
\footnotesize
\begin{tabular}{lcl}
\hline\hline\noalign{\smallskip}
Parameter
&Value
&Fixed/free parameter and comments
\\
\noalign{\smallskip}
\hline\noalign{\smallskip}
Disk central velocity (\kms)				&$-43.0$		&Fixed from observed PA=40\degr\ pv-plot\\
Disk outer radius ($''$, adopted as reference radius)	&$+5$		&Fixed from emission size\\
Disk inner radius ($''$)				&$+1$		&Fixed: upper limit from SMA data (see main text)\\
Intensity power law index			&$-1$		&Fixed following other studies of disk-like structures (see main text)\\
Infall velocity power law index				&$-0.5$		&Fixed, free-fall\\
Rotation velocity power law index			&$+1$		&Fixed, solid rigid rotation\\
\hline\noalign{\smallskip}
Reference infall velocity (\kms)  &$0.17\pm0.02$&Free: fitted in the PA=40\degr\ pv-plot\\
Reference rotation velocity (\kms)&$0.6\pm0.2$&Free: fitted in the PA=130\degr\ pv-plot, after fitting the ref. infall vel.\\
\hline
\end{tabular}
\label{tn2hpvinfrot}
\end{table*}

\begin{table*}[ht]
\caption{Physical parameters of the outflow driven by MM1}
\centering
\footnotesize
\begin{tabular}{lccccccccc}
\hline\hline\noalign{\smallskip}
&$t_\mathrm{dyn}$
&size
&$N$~$^\mathrm{a}$
&$M_\mathrm{out}$~$^\mathrm{a}$
&$\dot{M}$~$^\mathrm{a}$
&$P$~$^\mathrm{a}$
&$\dot{P}$~$^\mathrm{a}$
&$E_\mathrm{kin}$~$^\mathrm{a}$
&$L_\mathrm{mech}$~$^\mathrm{a}$
\\
Lobe
&(yr)
&(arcsec)
&(cm$^{-2}$)
&(\mo)
&(\mo~yr$^{-1}$)
&(\mo~\kms)
&(\mo~\kms~yr$^{-1}$)
&(erg)
&(\lo)
\\
\noalign{\smallskip}
\hline\noalign{\smallskip}
Red	&\phn3600	&$12\times9$	&$3.1\times10^{16}$	&\phn0.05	&$1.5\times10^{-5}$	&\phn1.2	&\phn$3.3\times10^{-4}$	&\phn$2.6\times10^{44}$	&0.46	\\
Blue	&\phn3300	&$14\times7$	&$2.9\times10^{16}$	&\phn0.05	&$1.4\times10^{-5}$	&\phn1.4	&\phn$4.2\times10^{-4}$	&\phn$4.2\times10^{44}$	&0.85	\\
All	&\phn3500	&$26\times9$	&$6.0\times10^{16}$	&\phn0.10	&$2.9\times10^{-5}$	&\phn2.6	&\phn$7.5\times10^{-4}$	&\phn$6.7\times10^{44}$	&1.31	\\
\hline
\end{tabular}
\begin{list}{}{}
\item[$^\mathrm{a}$] Parameters are calculated for an inclination with
  respect to the plane of the sky equal to 0\degr, and are
  corrected for opacity effects. $N$ refers to CO column density.
\end{list}
\label{toutfpar}
\end{table*}


\begin{figure}[hpt!]
\begin{center}
\begin{tabular}[b]{c}
        \epsfig{file=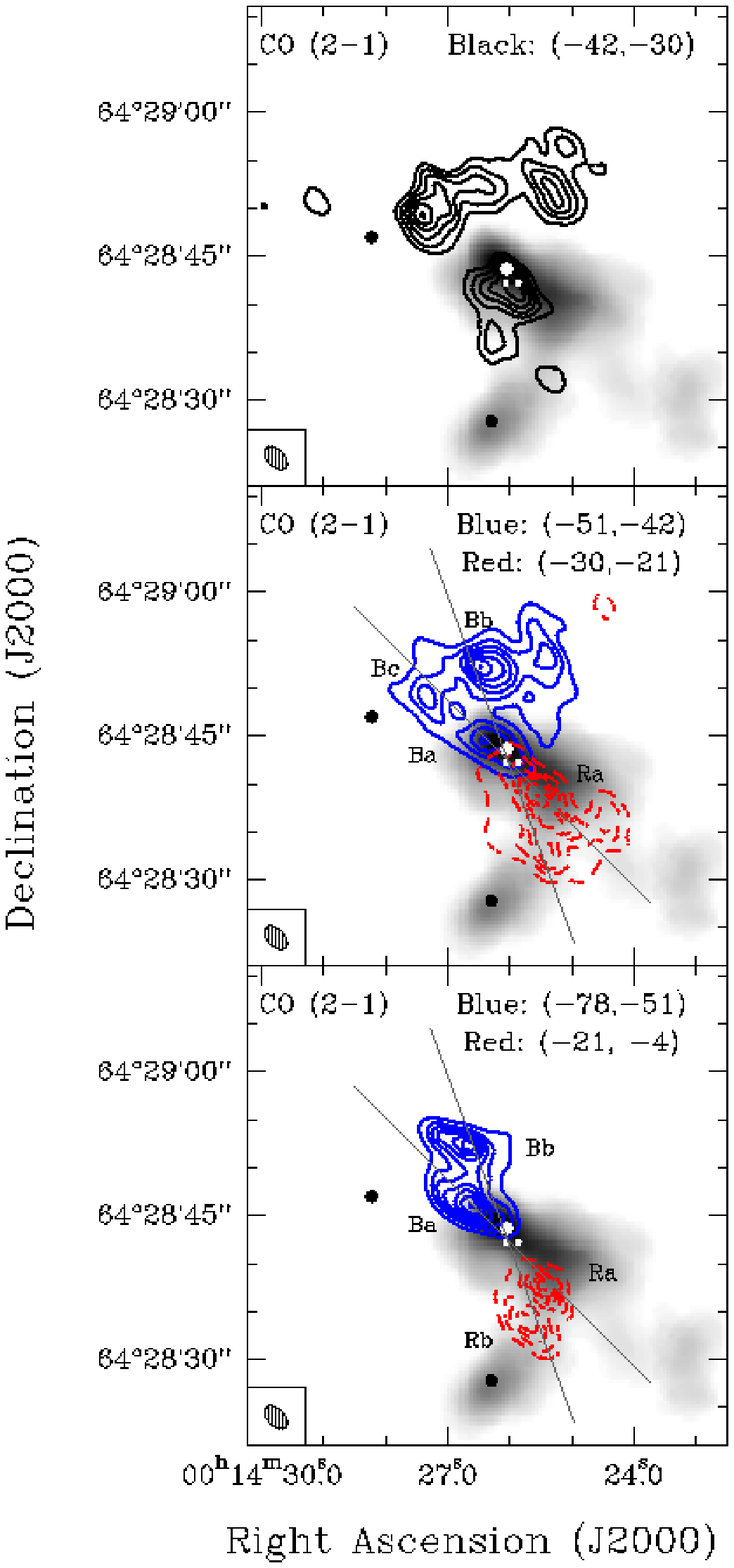, scale=0.69} \\
\noalign{\bigskip}
\end{tabular}
\caption{
\emph{Top}: Contours: CO\,(2--1) moment-zero map integrated for
systemic velocities (between $-6$ and $+6$~\kms\ with respect to the
systemic velocity, $-36.3$~\kms). Levels start at 24\%, increasing in
steps of 15\% of the peak intensity, 22.9~\jpb~km~s$^{-1}$.
\emph{Middle}: Idem for moderate velocities. Blue (solid) contours: emission
in the velocity range from $-6$ to $-15$~\kms\ with respect to the systemic
velocity. Red (dashed) contours: emission in the velocity range from $+6$ to
$+15$~\kms\ with respect to the systemic velocity. Levels start at 9\%,
increasing in steps of 15\% of the peak intensity,
32.7~\jpb~km~s$^{-1}$ and 25.8~\jpb~km~s$^{-1}$ for the blue and
red-shifted velocities, respectively.
\emph{Bottom}: Idem for high velocities. Blue (solid) contours: emission in
the velocity range from $-15$~\kms\ to most negative values with respect to
the systemic velocity. Red (dashed) contours: emission in the velocity range
from $+15$~\kms\ to most positive values with respect to the systemic
velocity. Levels start at 9\%, increasing in steps of 15\% of the peak
intensity, 13.0~\jpb~km~s$^{-1}$ and 11.1~\jpb~km~s$^{-1}$ for the blue
and red-shifted velocities, respectively.
In all panels, grey scale: \nth\ zero-order moment map; grey solid
lines indicate the possible orientation (PA=45\degr\ and 20\degr) of
the outflow(s). Black/white dots indicate the position of the \uchii\
region, MM1 (main, south, and southwest), and MM2. The synthesized
beam, $2\farcs92\times1\farcs78$ at $\rm{P.A.}=44\fdg7$, is shown in
the bottom left corner of each panel.}
\label{fi_12co_mom0}
\end{center}
\end{figure}

\subsection{\nth\ kinematics: MM2 \label{samm2}}

We fitted the \nth\,(1--0) spectrum toward MM2 with a single velocity
component (Fig.~\ref{fn2hspechfs}-bottom). As can be seen in the
figure, there is some excess of emission at blueshifted velocities,
which could be contamination from the MM1 ridge (which has an
extension toward the south at blueshifted velocitites with respect to
the MM2 clump, at around ($-44.6$)--($-44.7$)~\kms, as seen in
Fig.~\ref{fn2hch}). Alghough the excess of emission is quite clear, a
fit with two velocity components for the MM2 \nth\ hyperfine spectrum
could not be well determined\footnote{This is because the
  two-velocity-components hyperfine fit is strongly dependent on the
  adopted initial values. For the case of MM1 and the NE position of
  the MM1 ridge, this problem was solved by fitting two Gaussian to
  the line which is isolated, for which both velocity components are
  well detected. However, in the case of MM2, the velocity component
  at $-44.5$~\kms\ is barely detected in the isolated line (see
  Fig.~\ref{fn2hspechfs}-bottom), hindering the estimate of the
  initial values.}. From the parameters obtained from the fit, we
derived the excitation temperature and \nth\ column density as in the
previous section, and the values found for MM2 are very similar to the
values found for MM1.

In order to further constrain the \nth\ kinematics toward MM2, we
approached the flattened structure seen in \nth\ to a model of a
spatially infinitely thin disk seen edge-on, with a power-law
intensity as function of radius, consisting of a superposition of
optically thin rings undergoing infall and rotation (also described as
power-laws). We computed the synthetic pv-plots along the projected
major and minor axes of the disk, with angular and spectral
resolutions of $3.4''$ and $0.35$~\kms, respectively. The modeled
emission is shown in Fig.~\ref{fn2hpvinfrot}-bottom, and the
adopted/fitted parameters are listed in Table~\ref{tn2hpvinfrot}.

The disk structure is modeled with an inner and outer radius. The
outer radius is constrained from the \nth\ emission in the observed
pv-plot, and is also taken as the reference radius for the rotation
and infall power-laws. The intensity power-law index was adopted to be
$-1$, as used in other models of disk-like structures around
intermediate/high-mass YSOs (\eg\ Beltr\'an \et\ \cite{beltran2004}).
For the infall we assumed a free-fall infall velocity, and for the
rotation law we assumed a solid rigid rotation, as recent studies show
that these initial conditions are able to account for most
observational properties of star-forming cores (\eg\ Tscharnuter
\et\ \cite{tscharnuter2009}; Walch \et\ \cite{walch2009}; Zhilkin
\et\ \cite{zhilkin2009}).  The free parameters are the reference
infall velocity and the reference rotation velocity.


We constrained the upper limit of the inner radius from the SMA map at
1~mm, which has an angular resolution of $\sim\!2.5''$, and hence any
inner gap must have a radius $\lesssim1.2''$.  On the other hand, the
\nth\ data further constrain the inner radius to $0.8''$, as a smaller
inner radius clearly cannot reproduce the double peak in the
PA=130\degr\ pv-plot. We conclude that the inner radius covers the
range of 0.8--$1.2''$, and adopted the mean value of $\sim1''$. We
derived the reference infall velocity from the PA=40\degr\ pv-plot
(only affected by infall), of $0.17\pm0.02$~\kms.  We note that to
include infall in the model is necessary in order to reproduce the
double peak seen in the pv-plot at PA=40\degr.  Finally, we fitted the
rotation velocity from the PA=130\degr\ pv-plot (affected by both
infall and rotation). We note that in the PA=130\degr\ pv-plot the
velocity component at $\sim\!-43.5$ could be contaminated by the MM1
ridge, and hence we focused our fit on the two central peaks at around
$-43.0$~\kms. The resulting reference rotation velocity is
$0.6\pm0.2$~\kms. We also note that if we fit the data adopting an
inner radius of $0.8''$ or $1.2''$, the new parameters are similar,
within the uncertainties, to the parameters derived for an inner
radius of $1''$.



\begin{figure}[hpt!]
\begin{center}
\begin{tabular}[b]{c}
        \epsfig{file=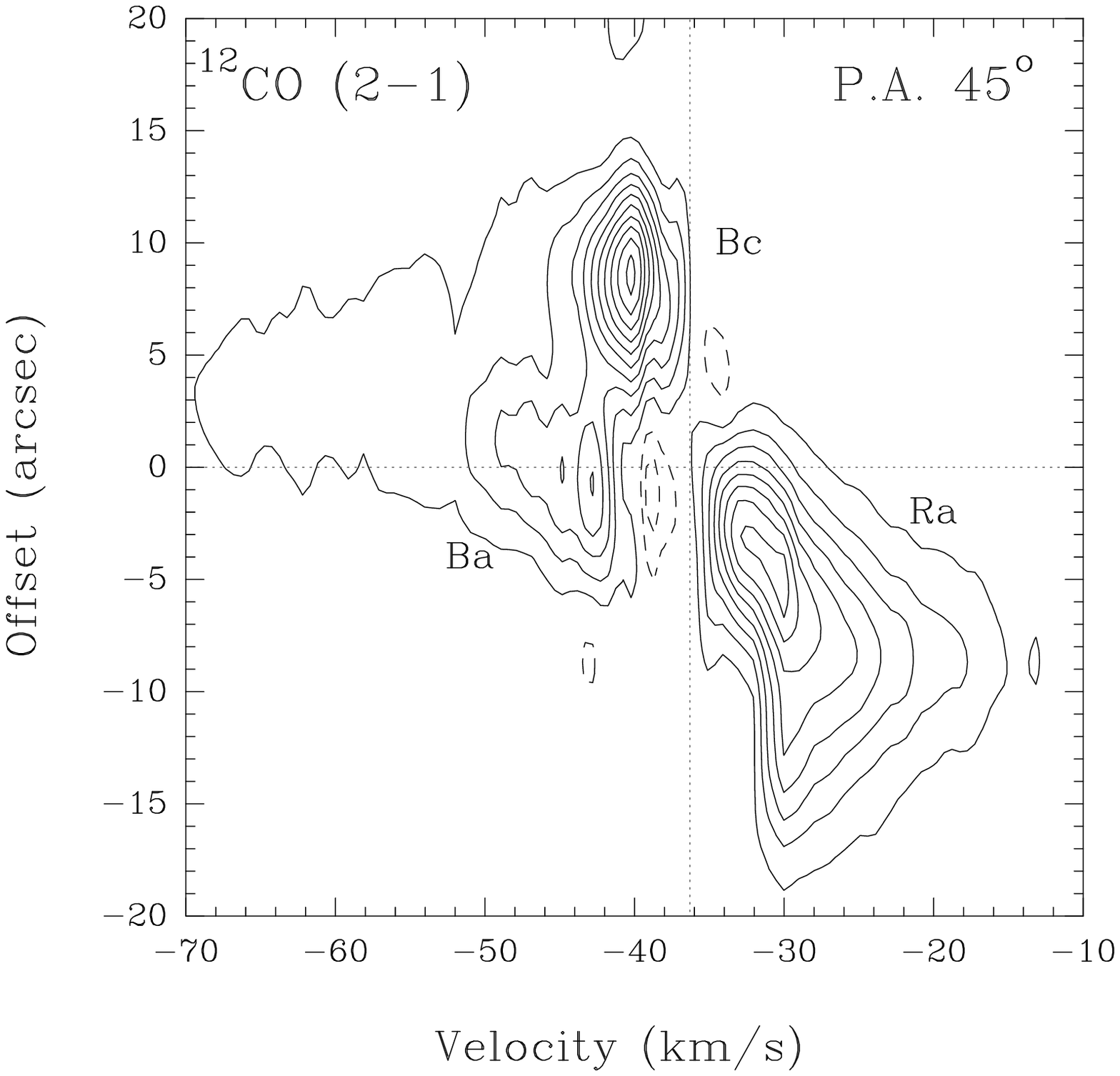, scale=0.47} \\
        \epsfig{file=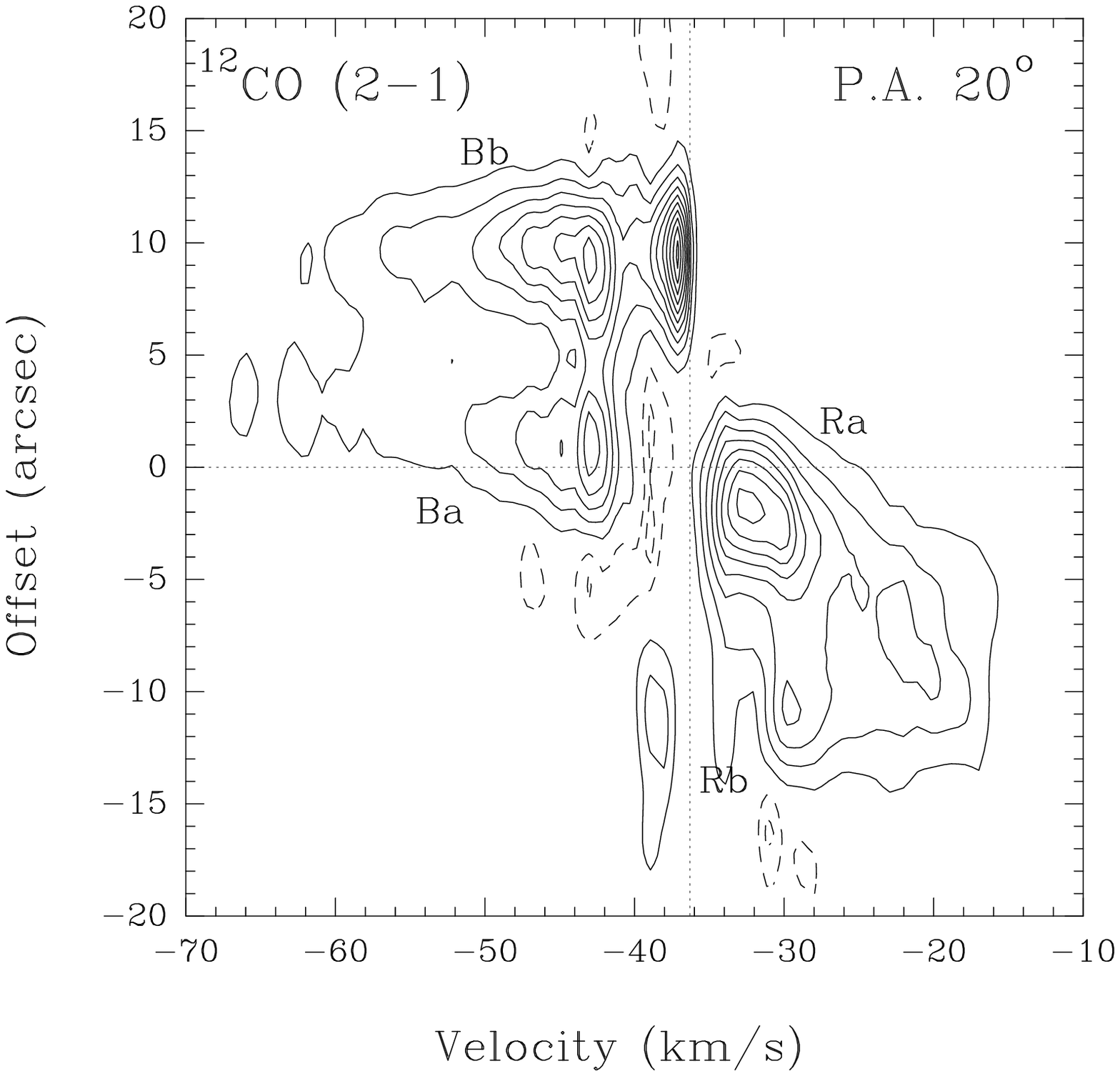, scale=0.47} \\
\noalign{\bigskip}
\end{tabular}
\caption{
\emph{Top}: CO\,(2--1) pv-plot in the northeast-southwest direction
with PA=45\degr, at the offset position ($3\farcs36$; $0\farcs96$), with
respect to the phase center.
\emph{Bottom}: CO\,(2--1) pv-plot in the north-south direction with
PA=20\degr, at the offset position ($2\farcs07$; $-0\farcs04$), with
respect to the phase center.
For both panels, contours are $-12$\%, $-6$\%, 6\% to 96\% in steps of
10\% of the peak intensity, 3.63~\jpb\ for the plot at PA=45\degr, and
3.59~\jpb\ for the plot at PA=20\degr.  Both pv-plots have been
convolved with a gaussian of $5\farcs0\times2\farcs0$, at
$\rm{P.A.}=135\fdg0$ and $\rm{P.A.}=110\fdg0$ respectively.
}
\label{fi_12co_pv}
\end{center}
\end{figure}

From the fitted infall velocity, one can estimate the mass of the
central object by using $M=v_\mathrm{inf}^2R/(2Gsin^2i)$, with $i$
being the inclination angle to the plane of the sky.  Thus, taking
$v_\mathrm{inf}\sim\!0.2$~\kms\ at a reference radius of $\sim\!5''$,
and assuming an inclination of 90\degr, we obtain a mass
$\sim>0.2$~\mo, where the lower limit accounts for the inclination
assumption.  We note that the true mass however is likely not larger
than a few tenths of solar mass, because we estimated the inclination
of the MM2 \nth\ core from its deconvolved major and minor axis
($8''\times1''$), and obtained an inclination, assuming it is
intrinsically circular, of $\sim80$\degr, which is very close to the
assumed inclination.
It is interesting to compare the mass derived from infall, which is
indicative of the mass already accreted onto the protostellar core,
with the mass derived from the millimeter continuum emission for MM2
(around 1.7~\mo, Sect.~\ref{srmm}), which is tracing the mass of the
envelope and disk surrounding the central core.  For Class 0 sources
it is typically assumed that these two masses should be of the same
order. Thus, the obtained accreted mass smaller than the disk+envelope
mass is suggesting that the central protostellar object in MM2 has
just recently formed.



\subsection{CO kinematics: moments and pv-plots}

We constructed the zero-order moment maps of the CO emission
integrated over three different velocity ranges: (1) systemic
velocities from $-6$ to $+6$~\kms\ with respect to the systemic
velocity (Fig.~\ref{fi_12co_mom0}-top), (2) moderate velocities from
$-6$ to $-15$~\kms, and from $+6$ to $+15$~\kms\ with respect to the
systemic velocity (Fig.~\ref{fi_12co_mom0}-middle), and (3) high
velocities from $-15$~\kms\ to most negative velocities, and from
$+15$~\kms\ to most positive velocities with respect to the systemic
velocity (Fig.~\ref{fi_12co_mom0}-bottom). At systemic velocities the
emission appears concentrated in two main structures, a clump
associated with the \nth\ MM1 ridge, and a multipeaked-arch located to
the north of MM1, just bordering the \nth\ emission.  At moderate and
high velocities, which are less affected by the missing short spacings
in the interferometric data, the emission has a bipolar structure
centered near the position of MM1 and is elongated in the
northeast-southwest direction (see Fig.~\ref{fi_12co_mom0}-middle and
bottom). At blue moderate velocities, the emission splits up into two
main peaks: Ba, close to the position of MM1 and elongated in the
northeast-southwest direction, and Bb, located $\sim\!10''$ northwards
of MM1 and with a round shape.  A third fainter clump, Bc, is located
$\sim 10''$ to the northeast of MM1.  Regarding the redshifted
emission, it appears as an elongated cone-like structure, with one
main clump, Ra, which is well aligned with the Ba clump, with a
position angle of $\sim\,45$\degr, and centered $\sim7''$ toward the
southwest of MM1.
At higher velocities, the emission is found toward Ba, Bb, Ra, and
toward a second redshifted clump, Rb, which was not very prominent at
moderate velocities. Clumps Bb and Rb are aligned at a position angle
of $\sim\,20$\degr, and centered near MM1. Note that the position
angles of the high-velocity CO emission are almost perpendicular to
the direction of the \nth\ velocity gradient, found at $\sim130$\degr.

\begin{table*}[ht!]
\caption{2MASS sources associated with the centimeter and/or millimeter emission toward the star-forming region IRAS~00117+6412}
\centering
\small
\begin{tabular}{lccccccccc}
\hline\hline\noalign{\smallskip}

&2MASS
&\multicolumn{2}{c}{Position}
&
&
&
&$J-H$
&$H-K_{\mathrm{S}}$
\\
Source
&Identification
&$\alpha (\rm J2000)$
&$\delta (\rm J2000)$
&$J$
&$H$
&$K_{\mathrm{S}}$
&color$^\mathrm{a}$
&color$^\mathrm{a}$
&IR$_\mathrm{excess}$$^\mathrm{b}$
\\
\hline\hline
\noalign{\smallskip}
\uchii\ 	&$00142828+6428471$	&00 14 28.29	&64 28 47.19
		&$14.94\pm0.08$	        &$13.28\pm0.07$	&$12.14\pm0.05$	 &$1.66$   &$1.14$  &$+0.18$	\\
MM1		&$00142616+6428444$	&00 14 26.16	&64 28 44.44	
		&$15.49\pm0.07$	        &$13.58\pm0.06$	&$12.22\pm0.04$	 &$1.91$   &$1.37$  &$+0.27$	\\
{\scriptsize 2M0014256}	&$00142558+6428416$	&00 14 25.59	&64 28 41.70 
                &$16.71\pm0.14$         &$14.35\pm0.05$ &$13.27\pm0.04$	 &$2.36$   &$1.09$  &$-0.27$	\\
\hline
\end{tabular}
\begin{list}{}{}
\item[$^\mathrm{a}$] Uncertainties in the colors are typically $\sim0.09$~mag.
\item[$^\mathrm{b}$] The infrared excess is measured as the difference
between the $(H-K_{\mathrm{S}})$ color and the $(H-K_{\mathrm{S}})$ color corresponding to a
reddened main-sequence star (following the reddening law of Rieke \&
Lebofsky \cite{riekelebofsky1985}), \ie\ $\mathrm{IR_{excess}}=(H-K_{\mathrm{S}})-0.576\times(J-K_{\mathrm{S}})$.
\end{list}
\label{t2mass}
\end{table*}


We performed different pv-plots close to MM1 at PA=45\degr, and
PA=20\degr (see Fig.~\ref{fi_12co_pv}), in the two directions
indicated in Fig.~\ref{fi_12co_mom0}. Both pv-plots have been convolved
with a Gaussian ($5''\times2''$ at PA=135\degr\ and 110\degr,
respectively). In both plots, positive and negative positions
correspond respectively to northeastern and southwestern positions
with respect to MM1. In the first cut (45\degr) we can distinguish a
strong velocity gradient from clump Ra to clump Ba, with velocities
increasing with distance, following a Hubble-law pattern (up to
34~\kms\ with respect to the systemic velocity in the case of clump
Ba). We note that in the cut at PA=45\degr, the clump Bc shows
moderate velocity emission of up to 14~\kms\ with respect to the
systemic velocity. Regarding the pv plot at PA=20\degr, 
the redshifted emission and the blueshifted clump Ba also follow a
Hubble-law, as in the cut at 45\degr.

We calculated the energetics of the outflow for each blue and red
lobe separately (for the blue lobe we included both clumps Ba and
Bb), assuming that all the emission comes from a single outflow, and
listed the values in Table~\ref{toutfpar}. The expressions used to
calculate the outflow CO column density, $N(\mathrm{CO})$, from the
transition $J \to J-1$, and the outflow mass, $M_\mathrm{out}$, are
given in Palau \et\ (\cite{palau2007b}). We adopted an opacity in the
line wings of $\sim2$ (from preliminary $^{13}$CO data), and an
excitation temperature of $\sim\,7$~K, estimated from the spectrum in
Fig.~\ref{fi_12co_spectra_1} (assuming that CO is optically thick and
adopting a line temperature of 2.7~K). For the red lobe we integrated
from $-30$ to $-8$~\kms, and for the blue lobe from $-72$ to
$-42.5$~\kms.  The dynamical timescale of the outflow,
$t_\mathrm{dyn}$, was derived by dividing the size of each lobe by the
maximum velocity reached in the outflow with respect to the systemic
velocity ($28.3$~\kms\ for the red lobe, and $35.7$~\kms\ for the blue
lobe).

\subsection{IRAS, MSX and 2MASS emission \label{ssed}}

With the aim of finding the possible infrared counterparts of the
sources studied in this work, we searched the IRAS, MSX and 2MASS
surveys (the region has not been observed by Spitzer). While the IRAS
position error ellipse makes it difficult to disentangle the
contribution from the \uchii\ region and MM1, we superposed the MSX
emission on our millimeter maps and found that the MSX emission is
compact and clearly peaking at the \uchii\ region. Thus, this is
suggestive of the major part of the IRAS and MSX fluxes, at least up
to $\sim20~\mu$m, coming from the \uchii\ region. However, at 60 and
100~$\mu$m the contribution of the \uchii\ region and MM1 is not
clear, and for this reason we refrained from building the spectral
energy distributions. As for the 2MASS Point Source Catalog (Skrutskie
\et\ \cite{skrutskie2006}), in Table~\ref{t2mass} we show the 2MASS
photometry for the 2MASS counterparts of the \uchii\ region, MM1, and
for 2M0014256. From the 2MASS magnitudes, we estimated the ($J-H$) and
($H-K_\mathrm{S}$) colors and measured the infrared excess as the
difference between the ($H-K_\mathrm{S}$) color and the
($H-K_\mathrm{S}$) color corresponding to a reddened main-sequence
star (following the reddening law of Rieke \& Lebofsky
\cite{riekelebofsky1985}). The 2MASS sources associated with the
\uchii\ region and MM1 show a moderate infrared excess typical of
Class~II sources, while 2M0014256 seems to be a reddened main-sequence
star, or a Class III source. A detailed analysis of the low-mass
content of the forming cluster in IRAS 00117+6412 will be presented in
a subsequent paper (Busquet \et, in prep.).

\section{Discussion \label{sdis}}

\subsection{A shell-like \uchii\ region}

\begin{figure}[t!]
\begin{center}
\begin{tabular}[b]{cc}
\noalign{\bigskip}
	\epsfig{file=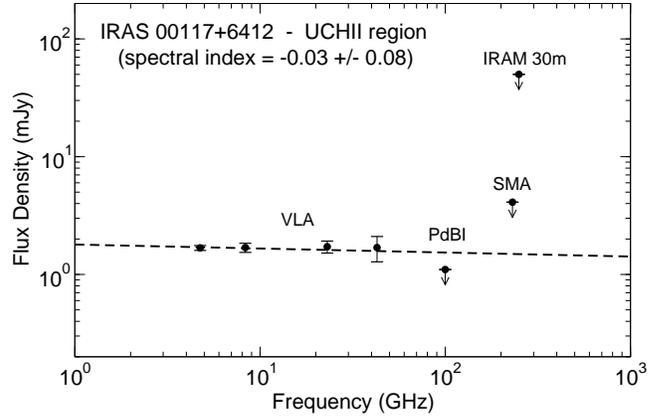, scale=0.32} \\
\noalign{\bigskip}
\end{tabular}
\caption{Spectral energy distribution in the centimeter and millimeter
  range for the \uchii\ region. Dashed line: free-free optically thin
  fit with a spectral index of $-0.03$. VLA, SMA and PdBI data from
  this work. IRAM~30\,m data from S\'anchez-Monge \et\ (\cite{sanchezmonge2008}).}
\label{fi_seds}
\end{center}
\end{figure}

The centimeter range (from 6~cm up to 7~mm) of the spectral energy
distribution of the \uchii\ region can be fitted assuming free-free
optically thin emission from ionized gas with a spectral index of
$-0.03\pm0.08$ (see Fig.~\ref{fi_seds}; the spectral index was
calculated from the fluxes measured in the images at 6, 3.6, 1.3~cm
and 7~mm made with the $uv$-range shared at all the wavelengths,
4.35--55~k$\lambda$).
We calculated the physical parameters of the
\hii\ region at 3.6~cm (in the CD configuration) assuming the emission
is optically thin. The results are consistent with a \uchii\ region
with a size of $0.03\pm0.01$~pc, a brightness temperature of 12.4~K,
an emission measure of $1.0\times10^{5}$~cm$^{-6}$~pc, an electron density
of 2000~cm$^{-3}$, a mass of inonized gas of
$4.2\times10^{-4}$~\mo\ (estimated from the beam averaged electron
density and the observed size of the source), and a flux of ionizing
photons of $3.8\times 10^{44}$~s$^{-1}$. These parameteres are
consistent with a \uchii\ region driven by an early-type B2 star
(from Panagia \cite{panagia1973}), as already stated by S\'anchez-Monge \et\
(\cite{sanchezmonge2008}).


The high angular resolution centimeter observations presented in this
work reveal that the \uchii\ region reported in S\'anchez-Monge
\et\ (\cite{sanchezmonge2008}) already has a shell-like structure. The
average radius of the shell (from an average of the three
subcondensations) is $\sim1\farcs2$, which corresponds to 2200~AU or
0.01~pc. From this radius, one can make a rough estimate of the
dynamical timescale, either assuming a classical expansion of an HII
region at $\sim10$~\kms, or the expansion of a wind-driven bubble (as
typically assumed to explain the shell-like morphologies of
\uchii\ regions, \eg\ Garay \& Lizano \cite{garaylizano1999}). In the
first case (classical HII region), the dynamical timescale is
$\sim1000$~yr, while in the second case (stellar wind blown bubble)
the dynamical timescale is $\sim 12000$~yr (assuming the stellar wind
dominates over the classical expansion of the \uchii\ region, an
initial ambient density of $\sim10^7$~\cmt, and following Castor
\et\ \cite{castor1975}, and Garay \& Lizano \cite{garaylizano1999},
yielding an expanding velocity $\lesssim1$~\kms). The lifetimes
derived from both assumptions are much smaller than $\sim1$~Myr, the
lifetime estimate for \uchii\ regions of $\sim 1000$~\lo, from the RMS
survey by Mottram \et\ 2009 (in prep.). This could be indicative of
the \uchii\ region in IRAS~00117+6412 having undergone in the past a
period of strong quenching (due to a high-mass accretion rate,
Walmsley \et\ \cite{walmsley1995}), before expanding as a classical
HII region or as a wind-blown bubble.  Alternatively, if the classical
HII region expansion or the wind-blown bubble assumptions are correct,
we could be witnessing the very first stages of the expansion of an
ionized shell around a B2-type ($\sim1000$~\lo) YSO. However, the fact
that we do not detect significant dust or dense gas emission
associated with the \uchii\ region suggests that the object is in the
process of disrupting its natal cloud, and most likely has a lifetime
similar to the $\lesssim1$~Myr given by Mottram et al. (2009, in
prep.). In addition, diffuse optical emission can be seen to the
northeast of the \uchii\ region in the blue and red plates of the
Palomar Observatory Sky Survey II, supporting the fact that the
\uchii\ region is not deeply embedded and is emerging from its natal
cloud.

From the bolometric luminosity and the spectral type of the
\uchii\ region one can estimate the mass and the evolution time from
the birthline by placing the object in a HR diagram. Following the
models of Palla \& Stahler (\cite{pallastahler1990},
\cite{pallastahler1993}), and adopting an effective temperature of
$\sim20500$~K (Panagia \cite{panagia1973}), the mass of the star
ionizing the \uchii\ region is about 5--6~\mo, and the star is already
placed at the ZAMS, which most likely is reached after $\sim1$~Myr for
a star of this mass (\eg\ Palla \& Stahler \cite{pallastahler1993};
Bernasconi \& Maeder \cite{bernasconimaeder1996}). This estimate 
agrees with the previous estimate of the lifetime of a
\uchii\ region of $\sim1000$~\lo\ (derived from Mottram \et\ 2009, in
prep.) of $\sim1$~Myr.

\subsection{MM1: multiple sources and a spectacular outflow}

In previous sections we showed that MM1 is associated with a 2MASS
source with infrared excess. Dense gas emission traced by \nth\ and
faint centimeter emission are also associated with MM1. These
properties suggest that MM1 is most likely in a phase of
moderate/intense accretion, possibly in the Class 0/I stage, with
typical ages of (0.5--1)$\times10^5$~yr (\eg\ Evans
\et\ \cite{evans2009}).
On the other hand, the estimated envelope mass of ~3.0~\mo\ and the
magnitude of the near-infrared emission (similar to the magnitudes of
the \uchii\ region, which has a luminosity of $\sim1000$~\lo) are
suggestive of MM1 being possibly an intermediate-mass YSO.
In particular, from the outflow momentum rate ($\dot{P}$) we can
estimate the luminosity of the object powering the outflow (assuming
one single outflow). Using the relation from Hatchell
\et\ (\cite{hatchell2007}) we obtain $L_\mathrm{bol}=400$~\lo, and
with the relation from Bontemps \et\ (\cite{bontemps1996}) we obtain
$L_\mathrm{bol}=560$~\lo, consistent with a driving source of
intermediate mass.

We note that concerning the driving source(s) of the outflow(s), we
discard 2M0014256 as a possible candidate because its near-infrared
excess is much smaller than that of MM1 (see Table~\ref{t2mass}), and
lies slightly off from the line joining the high-velocity clumps Ba
and Ra, or Bb and Rb. The only aspect suggestive of 2M0014256 being
associated with the star-forming region is the elongation seen at
3.2~mm, which matches well with the position of the 2MASS source (see
Fig.~\ref{fi_cont}e). However, such an elongation at 3.2~mm could be
tracing dust entrained by the outflow, as the elongation and the Ba
and Ra lobes are very well aligned. If 2M0014256 was really associated
with the region one would expect some compact emission associated with
it in the 1~mm continuum maps (which have higher angular resolution
than the 3.2~mm continuum maps), but as this is not found, it suggests that
the elongation is made of extended emission.

Another aspect favoring the scenario that MM1 could be the driving
source of the outflow is that MM1 is associated with centimeter
emission, typically found toward Class 0/I sources driving
outflows. Since we only detect the source in the centimeter range at
3.6~cm we can only estimate a range of possible spectral indices,
$-0.3<\alpha_{\rm{MM1}}<1.1$, which is consistent with thermal
free-free emission from ionized gas (the spectral index was calculated
from the images at 6, 3.6, and 1.3~cm made with the same
$uv$-range). The fact that the centimeter source is elongated in a
position angle of $\sim120\pm30$\degr, almost perpendicular to the CO
outflow, could be due to contribution from the different
subcondensations associated with MM1. Additionally, from the work of
Shirley \et\ (\cite{shirley2007}), we estimated the predicted
centimeter flux which can be accounted for from the outflow momentum
rate, and obtained a flux density at 3.6~cm of $\sim0.3$~mJy. Since we
measured a flux density at 3.6~cm of 0.17~mJy, we conclude that all
the centimeter emission associated with MM1 can be accounted for
through ionization by shocks from the outflow.

The interpretation of the high-velocity CO bipolar outflow and the
identification of the driving source(s) are not straightforward. A
possibility is that clumps Ba and Ra are tracing an outflow with a
position angle of 45\degr\ powered by one of the subcondensations of
MM1, maybe MM1-S, as it falls exactly at the center of symmetry of
clumps Ba and Ra (although it is less massive than MM1-main); and that
Bb and Rb are tracing a second outflow with a position angle of
20\degr, possibly powered by MM1-main, as it shows an elongated
structure in the east-west direction, almost perpendicular to the
Bb-Rb outflow.  Another possibility is that all the CO emission comes
from a single wide-angle outflow, which would be driven by MM1-main
(since it shows an east-west elongation), and would be excavating a
cavity, seen at moderate velocities. If this was true, the eastern
wall of the cavity, corresponding to clump Ba, would have very
high-velocity emission (up to $\sim35$~\kms\ with respect to the
systemic velocity). A similar case in the literature is found for
G240.31+0.07 (Qiu \et\ \cite{qiu2009}), a massive YSO driving a
wide-angle outflow, and with velocity in the walls of up to
27~\kms\ with respect to the systemic velocity.  Such a high velocity
in the cavity wall could be explained through precession of the
outflow axis and episodic mass ejection. Arce \& Goodman
(\cite{arcegoodman2001}) study the position-velocity relation for
episodic outflows and show that clumps at different positions from the
driving source are expected. Each one of these clumps shows a wide
range of velocities (see \eg\ Fig.~2 in Arce \& Goodman
\cite{arcegoodman2001}). This can also be seen in the pv-plot of
IRAS\,00117+6412 at 20\degr\ (Fig.~\ref{fi_12co_pv}-bottom).  In the
figure, both the blueshifted and the redshifted sides show a pair of
clumps, each one spanning a range of $\sim20$~\kms. Thus, both
possibilities seem plausible for this target, both mutiple outflows
driven by different subcondensations of MM1, and a single wide-angle
episodic outflow.


\subsection{MM2: an intriguing protostellar object}

In Sect.~\ref{samm2} we modeled the \nth\ emission toward MM2 as a
disk structure undergoing infall and rotation. The infall and rotation
velocities (at $5''$) can be compared with those derived from other
protostellar cores.  Beltr\'an \et\ (\cite{beltran2005}) model, for the
high-mass case for example, two massive cores with infall velocities
of 1--2~\kms, and rotation velocities of $\sim2$~\kms, at spatial
scales of 5000--10000~AU.  On the other hand, infall and rotation
velocities for low/intermediate-mass protostars are around 0.1--0.4
and 0.2~\kms, respectively, at spatial scales of $\sim 3000$~AU
(Beltr\'an \et\ \cite{beltran2004}; Ward-Thompson
\et\ \cite{wardthompson2007}; Carolan \et\ \cite{carolan2008}).  From
the infall and rotation velocities at the reference radius derived in
Sect.~\ref{samm2} for MM2, and using the assumed infall and rotation
power laws (Table~\ref{tn2hpvinfrot}), we found an infall and rotation
velocity of 0.29 and 0.2~\kms\ respectively at 3000~AU, which is
more similar to the velocities derived for low/intermediate-mass cores
than to the velocities derived for high-mass cores.
In addition, the flattened structure was modeled with an inner gap of
$\sim1''$. We note that this inner gap must not necessarily be a
physically real gap. First, it could be an opacity effect. 
However, this possiblity can be discarded because the opacity of the
isolated line (the hyperfine used to fit the model in
Sect.~\ref{samm2}) estimated from the fits is $\sim0.05$ (see
Table~\ref{tn2hh}), which is clearly optically thin.
Second, a more plausible option is that the \nth\ gap is reflecting
the depletion of \nth\ in the center, maybe due to freezing-out of
\nth\ due to high densities and low temperatures in the center, as
found in very young (sometimes pre-protostellar) and dense
($5\times10^5$~\cmt) regions (Belloche \& Andr\'e
\cite{bellocheandre2004}; Pagani \et\ \cite{pagani2007}). High angular
resolution observations of the continuum millimeter emission would
help assess the nature of this \nth\ gap.




It is worth noting that the properties of MM2 are quite different from
those of MM1.  MM2 has no infrared emission and is embedded in a
dusty compact condensation and in a dense gas clump, which seems to be
undergoing rotation and infall motions. As there is water maser
emission associated with the MM2 peak (see Fig.~\ref{fi_cont}f),
indicative of stellar activity, one would classify this source as
Class 0, judging from its lack of infrared emission and its strong and
compact millimeter continuum emission. However, YSOs in the Class 0
evolutionary stage are typically associated with strong and collimated
molecular outflows (\eg\ Bachiller \cite{bachiller1996}), while we
found no hints of (high-velocity) CO associated with MM2. Furthermore,
water masers associated with YSOs are related to accretion or ejection
processes (\eg\ Furuya \et\ \cite{furuya2003}), and observational
studies suggest a connection between water masers and the outflow
phenomenon (\eg\ Zhang \et\ \cite{zhang2001}). Thus, the lack of
outflow emission associated with MM2 is intriguing.  Different
possibilities could explain this behavior (lack of outflow in an
apparent Class 0 source with water maser emission). First, the object could be
driving an outflow very faint in CO, but which could be well traced by
other molecules such as SiO (studies of high-mass star-forming regions
show that some molecular outflows are faint or not detected in CO
while they are strong in SiO, and viceversa, Beuther et
al. \cite{beuther2004}; Zapata et al. \cite{zapata2006}). However, a
preliminary reduction of SiO\,(1--0) data observed with the VLA
(Busquet \et, in prep.) show no outflow hints. A second possibility
would be that the outflow has not been created yet (the protostellar
wind would still be in the phase of sweeping the ambient material
out). The caveat with this possibility is that observations seem to
indicate that outflows appear at the very first stages of the
protostar formation (\eg\ Bachiller \cite{bachiller1996}), and
the probability of witnessing such a short-lived phenomenon is
low. Third, the ejection process in this source could be different
from the standard paradigm of a strong and collimated outflow
associated with a Class 0 source. This is found for some objects in
massive star-forming regions (Torrelles \et\ \cite{torrelles2001},
\cite{torrelles2003}), which show spherical ejections traced by water
masers. Thus, the true evolutionary stage of MM2 remains an open
question. A study of the water maser emission with the highest angular
resolution available would be of great help to elucidate the process
of accretion/ejection in this enigmatic object.

In order to make a rough estimate of the MM2 luminosity we used the
water maser luminosity reported by Wouterloot
\et\ (\cite{wouterloot1993}) toward MM2, of $1.7\times10^{-6}$~\lo,
which we corrected to our adopted distance. From a correlation between
the water maser luminosity and the bolometric luminosity of the YSO
associated with the maser (Furuya \et\ \cite{furuya2003},
\cite{furuya2007}), we estimated a luminosity for MM2 of $\sim600$~\lo,
suggesting that MM2 is of intermediate-mass.


\subsection{Different objects emerging from the same natal cloud}

The results obtained toward IRAS~00117+6412 reveal that the dusty
cloud harbours about three intermediate-mass YSOs showing different
properties.  One of the sources seems to be an intermediate-mass
\uchii\ region with a shell-like structure, located at the eastern
border of the dusty cloud, and almost deprived from dust and dense
gas, with an estimated luminosity of $\sim1000$~\lo, and an estimated
age of $\sim1$~Myr. Another source, MM1, is deeply embedded within the
dusty cloud, has about 400--600~\lo, and an estimated age around
$\lesssim10^5$~yr. Thus, MM1 presumably formed \emph{after} the
\uchii\ region. Finally, MM2, of about $\sim600$~\lo, remains the most
enigmatic object in this region, as it seems to be deeply embedded in gas
and dust and has a water maser associated, but no signs of CO
outflow activity.  This could be
a new type of object undergoing a special process of matter ejection
(such as spherical mass ejection).

In summary, our observations show that the formation of stars within
the nascent cluster in IRAS\,00117+6412 seems to take place in
different episodes. In addition, these observations show that the
similar initial conditions within a cloud (similar dust mass,
excitation temperature, and dense gas column density, as found for MM1
and MM2) can yield objects with very different properties. This 
indicates that these initial conditions may not be decisive in
determining some of the properties of the YSOs forming within the
cloud, such as the ejection properties, a result already found at much
smaller spatial scales by Torrelles \et\ (\cite{torrelles2001},
\cite{torrelles2003}).


\section{Conclusions}

In this paper we study with high angular resolution the centimeter,
and millimeter continuum, and \nth\,(1--0), and CO\,(2--1) emission of
the intermediate-mass YSOs forming within a dusty cloud, with the goal
of assessing the role of the initial conditions in the star formation
process in clusters. Our conclusions can be summarized as follows:

\begin{enumerate}

\item A \uchii\ region is found at the eastern border of the dusty
  cloud, with a shell-like structure and a flat spectral index,
  $-0.03\pm0.08$. The estimated age and mass of the underlying star is
  $\sim1$~Myr and $\sim6$~\mo.

\item Deeply embedded within the dusty cloud, we have discovered a
  millimeter source, MM1, associated with a 2MASS infrared source,
  which is driving a CO\,(2--1) powerful and collimated high-velocity
  outflow, oriented in the southwest-northeast direction. The mass
  derived from the millimeter continuum emission for MM1 is
  $\sim3$~\mo. MM1 is embedded within a ridge of dense gas as traced
  by \nth, which seems to be rotating roughly along the outflow axis.
  MM1 is associated with centimeter emission, whose spectral index is
  compatible with an ionized wind, and at 1.2~mm splits up into
  different subcomponents when observed with an angular resolution of
  $\lesssim1''$. From the derived outflow momentum rate, we estimated
  a luminosity for MM1 of 400--600~\lo. Thus, MM1 seems to be a
  Class 0/I intermediate-mass YSO.

\item About $\sim15''$ to the south of MM1, our observations revealed
  a dust compact condensation, MM2, lying in a dark infrared region,
  associated with water maser emission and a dense core traced by
  \nth\ emission. The mass from the dust emission is $\sim1.7$~\mo,
  and the \nth\ excitation temperature and column density are similar
  to the ones derived for MM1. The dense core in MM2 is rotating in
  the same sense as the ridge associated with MM1 and seems to be
  undergoing infall. We modeled the MM2 dense core as a disk-like
  structure with an inner radius of $\sim1''$ and an outer radius of
  $\sim5''$, with a rotation velocity in the outer radius of
  $\sim0.6\pm0.2$~\kms, and an infall velocity at the same radius of
  $\sim0.17\pm0.02$~\kms. The non-detection of CO at any velocity
  toward MM2 makes this object intriguing.

\item Although MM1 and MM2 formed within the same cloud and have
  similar dust and dense gas emission, their properties, specially
  concerning the ejection phenomenon, seem to be different and could
  be indicating that the initial conditions in a cloud forming a
  cluster are not the only agent determining the properties of the
  members of the cluster.

\end{enumerate}

\begin{acknowledgements}

A. P. is grateful to Itziar de Gregorio-Monsalvo for useful
discussions.  A.P. is partially supported by the MICINN grant
ESP2007-65475-C02-02, the program ASTRID S0505/ESP-0361 from La
Comunidad de Madrid and the Europan Social Fund, and the Spanish
MICINN under the Consolider-Ingenio 2010 Program grant
CSD2006-00070. A.P., A.S.-M., G.B. and R.E. are supported by the
Spanish MICINN grant AYA2005-08523-C03, and the MICINN grant
AYA2008-06189-C03 (co-funded with FEDER funds).  This publication
makes use of data products from the Two Micron All Sky Survey, which
is a joint project of the University of Massachusetts and the Infrared
Processing and Analysis Center/California Institute of Technology,
funded by the National Aeronautics and Space Administration and the
National Science Foundation.

\end{acknowledgements}

{}

\begin{appendix}
\section{Emission filtered out by an interferometer \label{apen1}}

In this appendix we describe the estimation of the fraction of flux
filtered out by an interferometer. Consider a bidimensional Gaussian
source with the flux density $S_{\nu}$ and the half-power diameter
$D$. Its intensity, $I(x,y)$, can be expressed as
\begin{equation}
I(x,y)
=
\bigg[\frac{4\,\mathrm{ln(2)}}{\pi\,D^{2}}\bigg]\,
S_{\nu}\,
\mathrm{exp}\bigg[\frac{-4\,\mathrm{ln(2)}\,(x^{2}+y^{2})}{D^{2}}\bigg]
.
\label{eq_intensity}
\end{equation}

The visibility, $V(u,v)$, of the Gaussian source corresponds to the fourier
transform of the intensity (Eq.~\ref{eq_intensity}),
\begin{equation}
V(u,v)
=
S_{\nu}\,
\mathrm{exp}\bigg[\frac{-\mathrm{ln(2)}\,(u^{2}+v^{2})}{[(2\,\mathrm{ln(2)})/(\pi\,D)]^{2}}\bigg]
.
\label{eq_visibility}
\end{equation}
and thus, the half-power $(u,v)$ radius, $r_{1/2}$, is
\begin{equation}
r_{1/2}
=
\frac{2\,\mathrm{ln(2)}}{\pi\,D}
,
\label{eq_radius}
\end{equation}
or in practical units, $r_{1/2}$ in $\mathrm{k}\lambda$ and, $D$ in
$\mathrm{arcsec}$
\begin{equation}
\bigg[\frac{r_{1/2}}{\mathrm{k}\lambda}\bigg]\,
=\,
91.019\,
\bigg[\frac{D}{\mathrm{arcsec}}\bigg]^{-1}
.
\label{eq_radiusprac}
\end{equation}

From Eq.~\ref{eq_radiusprac} we can estimate the largest structure,
$\theta_\mathrm{LAS}$, sensitive to an interferometer with the shortest baseline
being $u_\mathrm{min}$,
\begin{equation}
\bigg[\frac{\theta_\mathrm{LAS}}{\mathrm{arcsec}}\bigg]\,
=\,
91.019\,
\bigg[\frac{u_{\mathrm{min}}}{\mathrm{k}\lambda}\bigg]^{-1}
.
\label{eq_thetalas}
\end{equation}

From Eq.~\ref{eq_visibility} and \ref{eq_thetalas} we can estimate the fraction
of correlated flux, $V(u_\mathrm{min})/S_{\nu}$, of a source with half-power
diameter $D$, when observed with an interferometer with a shortest baseline of
$u_\mathrm{min}$,
\begin{equation}
\frac{V(u_\mathrm{min})}{S_{\nu}}\,
=\,
\mathrm{exp}\bigg[-\mathrm{ln(2)}\,\bigg(\frac{u_\mathrm{min}\,D}{91.019}\bigg)^{2}\bigg]
,
\label{eq_fraction}
\end{equation}
with $u_\mathrm{min}$ in $\mathrm{k}\lambda$ and $D$ in $\mathrm{arcsec}$.

\end{appendix}






\end{document}